# A novel *agile* THz Pulsed Spectropolarimeter measuring 2D distributions of the magnetic field, density and ion temperature of fusion reactor equilibria

Roger J. Smith

1121 Lupine, Lake Forest, California 92630, *USA*

With the recent inception of intense pulsed THz sources via optical rectification using crystalline lithium niobate, non-perturbative measurements of the internal '*local*' magnetic field are now possible using Pulsed Polarimetry on any magnetic fusion device. Nowhere is the need for internal measurements more acute or the diagnostic so efficacious than for the FRC equilibrium. In particular, the amount of reversed (trapped) poloidal flux and the poloidal flux function are measurable for the first time. Pulsed Polarimetry resolves the sightline magnetic field and density, analogous to RADAR measurements of the spatial distribution of precipitation. With a suitable sightline, the FRC midplane $n_e(r)$ and $B_z(r)$ profiles can be determined with the azimuthal current density, $j_\varphi(r)$ given by Ampere's law. Pulsed Polarimetry can be enhanced to measure the sightline *distributed* electron temperature, $T_e(s)$ by adding spectroscopy. A THz Pulsed Spectropolarimeter using a heterodyne receiver is appropriate if collective Thomson backscatter is induced. In this case, electronic filtering provides spatial distributions of the ion spectral density function. The sightline bulk ion temperature, $T_q(s)$ as well as confined fast ion velocity distributions can be measured which is of particular importance to plasmas that are heated and sustained by NBI. The ability to steer the sightline without loss of alignment distinguishes this diagnostic from crossbeam diagnostics. The term, *diagnostic agility* denotes a diagnostic's ability to provide both temporal and *spatial* real-time feedback *on-demand* across the poloidal plane analogous to 2D imaging Doppler RADAR. THz Pulsed Spectropolarimetry is extensively detailed along with simulated measurements on an FRC equilibrium. Comparisons are made with long pulse CTS crossbeam diagnostics. The role this technique can play within the wider MFE program towards future fusion reactors as an *advanced* diagnostic is explored. The nature and breadth of the measurements fulfills the mission of future diagnostics to maintain and optimize fusion reactor performance. US patent #12,270,707.

## I INTRODUCTION

Diagnostics choices for fusion reactors will be based on their performance in maintaining operational control and optimization, without which burning plasmas would not be functional. The diagnostic technique of THz Pulsed Spectropolarimetry contributes *real-time* internal measurements of *key* equilibrium plasma parameters and is able to measure energy distributions of confined fast ions and alphas. The diagnostic is appropriate for all MFE equilibria but of special interest due to the absence of an equilibrium toroidal field are FRC plasmas. A first application of Pulsed Spectropolarimetry to FRCs would demonstrate the diagnostic's strengths while making seminal measurements on an equilibrium that has confounded our field for over 60 years.

The FRC is a highly attractive MHD configuration for a fusion reactor. The FRC community is making great strides in demonstrating FRC formation and sustainment by means of NBI and end biasing[1,2,3] instead of the traditional and technologically challenging theta pinch formation based on magnetic reconnection. FRC equilibria avoid the micro-instabilities of tokamaks[4] and are a pathway to advanced aneutronic fusion using p-B$^{11}$[5]. FRCs are stable to rotational instabilities by external biasing and demonstrate >40*ms* lifetimes, sufficient for active feedback, fueling by compact toroid injection and plasma shaping with programmable coils.

The FRC equilibrium[6] is an enigma even after 60 years of research. Stability is unexplained. At O and X points($B_{pol}$=**0**) gyrokinetic theory doesn't apply. FRCs are high <$\beta$> with large density/field gradients and a *self-organized loop voltage*. What are not known are the internal details of the equilibrium, notably the amount of reversed poloidal flux generated in formation and its history with NBI and end biasing. A diagnostic to directly measure, nonperturbatively, the internal magnetic field has been lacking. Diagnostics fail due to the aggressive nature of the FRC equilibrium, the same that attracts the plasma community: little to no toroidal field and consequently very high $<\beta>$~90%: high density together with low $|\boldsymbol{B}|$. Traditional magnetic field sensitive techniques such as X-mode reflectometry and multi-chord polarimetry are difficult to apply and, in some cases, as for the MSE diagnostic, ineffective. A THz Pulsed Spectropolarimeter is up to the challenge and well matched to the FRC. Highly elongated FRCs are stable, allowing good sightline access to midplane parameters. Achieving fusion with only low $<\beta>$ plasmas would be a shame if 20 years from now, the MFE community had not fully committed to understanding the FRC equilibrium with $<\beta>$ 100x that of the tokamak equilibrium.

Pulsed Polarimetry, a LIDAR–*like* technique, has for many years awaited the recent generation of a high energy ultrashort THz pulsed source. One can then 'build out' an appropriate THz source. The wavelength range, $\lambda$=0.1-3*mm*, is a confluence, or sweet spot, for polarimetry and collective Thomson scattering diagnostic techniques of utility for fusion reactors. *Local* measurements (*distributions*) of density, magnetic field and electron/ion spectral features (both bulk and confined fast ion) are obtained along arbitrary sightlines suggesting a new concept, *diagnostic agility*. The THz PP technique applies to all MHD equilibria: FRC, stellarator, tokamak, RFP and spheromak with peak $B_{pol}$ ≥0.5T, $n_e$≥3x10$^{19}$$m^{-3}$ and $L_p$≥0.5*m*, essentially all devices within the MFE program. The technique improves with increasing peak $n_e$, $B_{pol}$ and $L_p$.

This paper is organized as follows: Pulsed Polarimetry is reviewed in **SECTION II**; in **SECTION III**, Pulsed Spectropolarimetry is introduced in two varieties, **IIIa)** direct detection and **IIIb)** heterodyne(het) receivers with instrumental comparisons to long pulse-high power CTS diagnostics; in **SECTION IV** measurement signal-to-noise for both ITS and CTS PS is given and used to qualify three THz PS scenarios, comparisons with LP-HP CTS is also discussed; the THz pulsed source and pulse conditioning is detailed in **SECTION V**; **SECTION VI** proposes a CTS PS diagnostic layout for an FRC equilibrium measuring midplane $n_e(r)$ and $B_z(r)$; in **SECTION VII** heterodyne THz PP and CW polarimetry are compared; **SECTION VIII** covers heterodyne THz PS as an *advanced* diagnostic for fusion reactors and motivates the term *diagnostic agility* with a conclusion given in **SECTION IX**. MKS units are used and a guide to notation and terminology is given in **APPENDIX A**. **APPENDIX B** details the magnetic geometry and the Mueller–Stokes' polarimetry method.

**II PULSED POLARIMETRY**[7] exploits the SOP of TS backscatter induced by a polarized light pulse transiting a magnetized plasma in the poloidal plane to measure the sightline distributions of $B_{\parallel}(s)$ and $n_e(s)$. The geometry is $\theta_{sc}$=$\pi$, $\boldsymbol{k}_{sc}$= –$\boldsymbol{k}_o$ and $\boldsymbol{k}_s$= $\boldsymbol{k}_o$–$\boldsymbol{k}_{sc}$ =4$\pi/\lambda_o\boldsymbol{\hat{s}}$. Intensities, $I_{s,p}(t)$ and $n_e(s)$=$C_n\sum_{j=s,p} I_j(t)$, are measured, $s(=ct/2)$ is the *time–of–flight* distance. $\Delta E_{sc}$ and $P_{sc}$ are given by (1a,b).

$$\Delta E_{sc}(s) = E_p n_e(s) r_e^2 \Delta\Omega(s)\Delta s \ [J] \quad (1a)$$
$$P_{sc} = \Delta E_{sc}/\tau_{\Delta s} = E_p n_e(s) r_e^2 \Delta\Omega(s) c/2 \ [W] \quad (1b)$$

$P_{sc}$=0.38·$E_{400}\Delta\Omega_{/10}n_{20}[\mu W]$. In cold plasma, Appleton-Hartree dispersion theory[8], when $\boldsymbol{B}_{pol}\cdot\boldsymbol{\hat{s}}$≠0, a circular birefringence is present(Faraday effect) which rotates $\psi$ of $\boldsymbol{E}$ by $\Delta\psi$=½($N_+$–$N_-$)$k_o\Delta s$ as given in Eq (2).

$$\Delta\psi = \omega_{pe}^2 \omega_{ce}/(2c\omega^2)\Delta s cos(\theta) = C_{FR}\lambda^2 n_e \boldsymbol{B} \cdot \Delta\boldsymbol{s} \quad (2)$$

The detected progressive integration of $\Delta\psi$ to $s$ is $\alpha_{tot}(s)$, Eq (3a), inverting (3a) gives $B_{\parallel}(s)$, Eq (3b). The rate of rotation, $\Delta\alpha_{tot}/\Delta s(s)$ and an independently measured $n_e(s)$ provide $B_{\parallel}(s)$. Field reversal occurs where $\Delta\alpha_{tot}/\Delta s(s)$=0. $B_{\parallel}(s)$ is *not* measured directly.

$$\alpha_{tot}(s) = 2 \cdot C_{FR} \cdot \lambda_o^2 \int_o^s n_e \boldsymbol{B}(s') \cdot d\boldsymbol{s}' [rad] \quad (3a)$$
$$B_{\parallel}(s) = \boldsymbol{B} \cdot \boldsymbol{\hat{s}} \cong \frac{1.9 \cdot 10^{12}}{\lambda^2 n_e(s)}\frac{\Delta\alpha_{tot}}{\Delta s}(s)[T] \quad (3b)$$

For LIDAR, $\alpha_{tot}(s)$=$\alpha_o(s)$+$\alpha_{sc}(s)$: a $\alpha_o(s)$ at $s$ and an emission $\alpha_{sc}(s)$ from $s$ back to the receiver. By non-reciprocity, $\alpha_{tot}(s)$=2$\alpha_o(s)$, not strictly true due to the $\lambda^2$ dependence of $\alpha_{sc}(s)$. The detected $\alpha_{tot}(s)$~$atan\sqrt{I_p/I_s}$, unwrapped. PP is essentially a scanning double-pass polarimeter with end mirror and light pulse identified. LIDAR allows a spatial dissection of $\alpha_{tot}(L_p)$.

A plasma Verdet constant is $V_{\lambda,n}$=15.1·$\lambda_{mm}^2$ $n_{20}$[°/T-cm] and $\Delta\alpha$=20$V_{\lambda,n}$·$B_{\parallel}$·$\Delta s_{10}$[°]. $\Delta\alpha(L_P)$ should be ~50°

The $\Delta s$ is given by the convolved $\tau_p$ and $\tau$ time responses, (4). LIDAR, being a temporal-spatial technique: a small $\Delta s$ implies a short $\tau_p$ (wide $\delta v_p(=1/\tau_p)$).

$$\Delta s = c/2 \sqrt{\tau^2 + \tau_p^2} \quad (4)$$

Setting $\tau$=$\tau_p$, defines $\tau_{\Delta s}$≡2$\Delta s/c$=√2$\tau$. For all measurements, the spatial resolution, $\Delta s$~$c\tau_{\Delta s}$/2, frequency resolution, $\Delta f$~$1/\tau_{\Delta s}$ and $SNR$~√$\tau_{\Delta s}$.

The probe frequency, $f_o(=\omega/2\pi)$ is typically well above $f_{pe}(=\omega_{pe}/2\pi)$=90√$n_{20}$[GHz] but refraction cannot be ignored. Refraction bends the sightline without misalignment as refraction is reciprocal, allowing a close



approach to cutoffs and a more relaxed regard towards density gradients.

Pulsed polarimetry combines Thomson scattering with polarimetry, two well developed and solidly understood diagnostic disciplines, the THz regime included. Two 'device–plasma' figures of merit apply to PP: [$n_eL_p$] product for TS and [$n_eL_pB_{pol}$] product for polarimetry as the signal strength improves with higher $n_e$, a higher [$n_eB_{pol}$] yields stronger magneto-optic activity and a larger device size, $L_p$, allows a larger $\Delta s$ at fixed $\Delta s/L_p$. Both FMs improve on fusion reactors, and the technology simplifies with larger $L_p$.

**III PULSED SPECTROPOLARIMETRY** adds spectroscopy. The $\mathcal{I}_{s,p}(\nu)(t)$ and the SDF of an ion species, charge $q$, $S_{q,e}(\nu)(t) \propto \sum_{j=s,p} \mathcal{I}_j / \sum_{j=s,p} I_j$ are measured. Two versions of PS: **IIIa)** ITS PS senses the *local* electron SDF, $S_e(\hat{s},\nu)(s)$ using ITS and **IIIb)** CTS PS uses a heterodyne receiver to sense the *local* ion SDF, $S_q(\hat{s},\nu)(s)$ (thermal and confined fast ions) using CTS. Both can measure *local* $n_e$ and $B_\parallel$. The great potential of Pulsed Spectropolarimetry is realized as $S_{q,e}(\nu)$ is directly proportional to the *distributed* $\hat{s}$ directed 1D velocity distribution functions, $Z_q^2 f_{q,e}(\hat{s},\nu)$, a kinetic feature of the plasma. $Z_q$ is 1 for H, D ions and 2 for alphas. *Local* measurements of $Z_q^2 f_{q,e}(\hat{s},\nu)(s)$ give unprecedented spatial coverage of the electron or ion microstate. The LIDAR TS diagnostic on the JET tokamak provided $n_e(s)$ and $S_e(\hat{s},\nu)(s)$[9].

The SDF from CTS, $S_{CTS}(\nu)$ is associated with ion motion. The electron produced backscatter is correlated with the ions when the Salpeter parameter $\alpha_s(\theta_{sc})=1/k_s\lambda_{De}>1$, Eq (5), with $\lambda_{De}= 74\mu m\sqrt{T_{e,10}/n_{20}}$.

$$\alpha_s(\theta_{sc}) = 1.06 \cdot \lambda_{mm}/\sin(\theta_{sc}/2) \sqrt{n_{20}/T_{e,10}} \quad (5)$$

The $\delta\nu_{CTS}$ for a Maxwellian distributed singly charged ion with $\alpha_s(\theta_{sc})>2$, $T_e=T_q$, and neglecting magnetic effects is given by Eq (6),

$$\delta\nu_{CTS} = 3.92 \cdot \sin(\theta_{sc}/2)/\lambda_{mm}\sqrt{T_{q,10}/M_{q,amu}}[GHz] \quad (6)$$

with $\int S_q(\hat{s},\nu)d\nu$ approaching ½ for large $\alpha_s(\theta_{sc})$ due to an additional self-shielding by ions.

Backscatter geometry(LIDAR) produces the widest $\delta\nu$ for a given $\lambda$, but $\delta\nu_{CTS}$ still requires a heterodyne receiver to resolve $S_{CTS}(\nu)$ where the CTS emission is superimposed on a mixer together with a strong(~1$m$W), usually electronic LO. Heterodyning faithfully preserves the TS emission spectrum as an electric field amplitude, $E_{TS}$ factor in the detected power, $P_{het}\sim E_{LO}\cdot E_{TS}$. $P_{het}$ is substantially amplified by $E_{LO}$ and the SDF is shifted to $\delta f_{CTS}=(\delta\nu-f_{LO})$ to near baseband, $f=0$ where electronic filtering or a wide-BW DAQ can resolve $\delta f_{CTS}$. $\delta f_{CTS}$ is partitioned into frequency bins electronically. For $N_f=6$, $\Delta f=1$GHz with $SNR_{\Delta f}$ roughly 60% of the full BW $SNR$. Sub-GHz $\Delta f$ only makes sense if $\tau_{\Delta s}>1ns$, as for fusion reactor devices. Fast ions injected at $E_{FI}=80–120$keV widen $\delta\nu_{CTS}$ and are better served by CTS PS.

**IIIa)** The $\delta\nu_{ITS}$ for THz ITS IPS is very wide, Eq (7). Receivers are usually DD.

$$\delta\nu_{ITS} = 170 \cdot \sin(\theta_{sc}/2)/\lambda_{mm}\sqrt{T_{e,10}}[GHz] \quad (7)$$

The wide $\delta\nu_{ITS}$ and $\lambda^2$ dependence in $\alpha_{sc}(s)$ imply $\alpha_{sc}(s)>\alpha_o(s)$. The polarimeter can include a filter at $\lambda_o$, width $\delta\lambda_o$ to contain this effect. The filtered intensities, $I_{p,s}=\langle \mathcal{I}_{p,s}\rangle_{\delta\lambda_o}$ produce a rotation $\alpha_o(s)+\langle\alpha_{sc}(\lambda,s)\rangle_{\delta\lambda_o} \cong 2\alpha_o(s)$ To demonstrate, non-relativistic $S_e(\lambda,T_e)$[10] are calculated and filtered to quantify this effect, see **Fig. 1**.

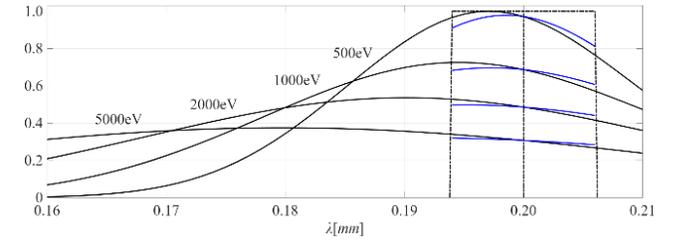

**Fig. 1.** Normalized $S_e(\lambda,T_e)$ at $T_e=\{0.5,1,2,5\}$keV, $\lambda_o=0.2mm$. The filter BW, $\delta\lambda_o$ is 12$\mu m$(90GHz). Blue curves are $\lambda^2/\lambda_o^2 S_e(\lambda,T_e)$ over the filter window (rectangle).

The integrals: $\int(\lambda^2/\lambda_o^2)S_e(\lambda,T_e)d\lambda \cong \delta\lambda_o\cdot S_e(\lambda_o,T_e) +$ *higher terms* and $\int S_e(\lambda,T_e)d\lambda$ over $[-\delta\lambda_o/2,\delta\lambda_o/2]$ are nearly identical, $\langle\alpha_{sc}(\lambda,s)\rangle_{\delta\lambda_o}\cong\alpha_o(s)$ to <0.3% at each $T_e$. The $\delta\lambda_o$ then, can be larger and is set by detector limits not by considerations of accuracy. Additional filter/detectors are used for measuring $T_e(s)$ and $n_e(s)$.

THz optical systems are characterized by $M(\cong 2A_d\Delta\Omega/\lambda_o^2)$, 2 polarization states. Large $\mathcal{E}$ optics are used for ITS PS, but aberrations: vignetting, a finite depth of field and an $\mathcal{E}\sim 1/s^2$, are introduced. For an $f^{10}$ telescope, $\Delta\Omega_{f/10}(s)$ at the focus is 0.008$sr$, $A_{4cm}(=12.6cm^2)$, $\lambda_o=0.2mm$ then $M=485$. The $SNR$ improves by $\sqrt{M}\sim 22$, over het detection, $M=1$, due to a *multiplex advantage*. Single shot measurements are possible with ITS PS.

**IIIb)** Het THz Pulsed Spectropolarimetry using CTS is the focus of this paper as it best fits NBI sustained FRCs but also for several remarkable properties of value to future fusion reactors, see **SECTIONS VII, VIII** and **IX**.

The origin of laboratory TS diagnostics is ionospheric sounding using RADAR in 1958, Bowles[11], resulting in



the discovery of collective Thomson scattering. The emission had an unexpectedly narrow $\delta v$ with, fortuitously, higher spectral brightness over the anticipated incoherent TS backscatter. Detection of ITS was demanding for RADAR sources and receivers at the time. The $\lambda_o$=7.3$m$ receiver was not heterodyne but a telescope for higher detected power.

The THz pulsed source that makes THz LIDAR possible, **SECTION V**, is based on converting NIR ultrashort pulses to THz pulses with $\eta$, presently, 2-4% at best. One might expect this loss to be defeating but the high spectral brightness of CTS combined with the exceptionally high $P_o$ of LIDAR redeems the technique. A 20$mJ$ laser produces a 400$\mu J$ over a 0.5$ns$ THz pulse with $P_o$=0.8MW comparing favorably to gyrotrons.

For a heterodyne PS, the launch, $\mathcal{L}$ and receive, $\mathcal{R}$ beams are *identified* Gaussian single mode, diffraction limited beams. Refraction is reciprocal, if $\mathcal{L}$ is refracted, the refraction is undone in $\mathcal{R}$. $\mathcal{E}=\lambda_o^2 [m^2 sr]$ for a single mode and doesn't vary with $s$. The spectral versions of $\Delta E_{sc}$, $P_{sc}$ Eqs (1a,b) become $\Delta\mathcal{E}_{sc}$, $\wp_{sc}$, Eqs (8a,b).

$$\Delta\mathcal{E}_{sc}(s) = E_p n_e r_e^2 \lambda_o^2 \Delta s / A_d S_q(v,s) \; [J/Hz] \quad (8a)$$

$$\wp_{sc} = \Delta\mathcal{E}_{sc}/\tau_{\Delta s} = E_p n_e r_e^2 \lambda_o^2 c / 2A_d S_q(v,s) \; [W/Hz] \quad (8b)$$

Eq 8b is the Thomson scattering *Equation of Transfer* to be compared to the CTS literature[12] Eq (9),

$$\wp_{sc} = (G) P_i O_b n_e r_e^2 \lambda_o^2 S_q(v) \quad (9)$$

$O_b \equiv V/A_\mathcal{L} A_\mathcal{R}$, possibly much less if refraction is present for a crossbeam system. Eqs 8b and 9 agree as $V(s)$=$A_d \Delta s$ and $P_i$=$E_p/\tau_{\Delta s}$ for LIDAR. $\wp_{sc}$=150·$E_{400} \lambda_{mm}^2 n_{20}/A_{1cm} S_q(v)$ [$nW$/Hz].

The fluctuations that produce the emission are detailed in $S_{TS}(v)$, the *dynamical form factor*. There is also a *geometrical form factor*, $G$ describing the coupling between incident and scattered EM fields when the dielectric properties of the medium are important. $G(V, k_o, \hat{e}_i, k_{sc}, \hat{e}_{sc})$[13,14] is the coupling coefficient of the polarization eigenstates, $\hat{e}_{i,sc}$ of the $\mathcal{L}$ and $\mathcal{R}$ beams given the eigenmodes of the plasma in V($s$). For CTS PS, $k_o$=−$k_{sc}$, $\mathcal{L}$ and $\mathcal{R}$ beams are same and $\hat{s}$ is arbitrary. For $\hat{s} \cdot B \neq 0$, the eigenstates are taken to be $\hat{e}_{+/-}$ and $G$ expresses the Faraday effect. For $\lambda$ far from cutoffs, $G$=1. LP-HP CTS diagnostics require the beams to be nearly perpendicular to magnetic flux surfaces to minimize refraction which can significantly lower $O_b$. In this case, the eigenstates are $\hat{e}_{O,X}$. $G$ can differ significantly from 1, even demonstrating ES($G$>>1), in particular, X–X scattering near cutoff[15]. A $\mu$wave LIDAR scattering diagnostic has been suggested for ES involving ions coupled to lower hybrid waves[16].

LP-HP CTS incurs unquantifiable measurement uncertainties because of $O_b$. Active feedback is required to keep $O_b$ at maximum with a subsequent increase in $\tau_{plasma}$. CTS PS remains aligned and can explore a larger parameter space regardless of refraction where CTS measurements may be of the greatest importance.

In LP-HP CTS, the central spectrum *must* be notched to block stray laser light. Beam/viewing dumps may be necessary, constraining the geometry further. LIDAR also has stray light issues in the form of pre-pulse and laser ASE, but THz LIDAR is *immune* to ASE due to the wavelength separation of the NIR and THz sources. For CTS PS, the entire SDF is available as any stray light is received after the scan. The central spectrum shape, if resolved, can express the *local* D/T:He ratios, important in fusion reactor operation/control. Since dumps are not required, *unrestrained steering* of the sightline is allowed. A LP-HP CTS diagnostic's main advantage is that $\alpha_s(\theta_{sc})$>1 can be more easily satisfied.

**IV MEASUREMENT SNR** calculations for a THz pulsed polarimeter technique were anticipated[17]. The photon noise is Gaussian distributed thermal noise due to an average mode occupancy, $n$, <# of photons per EM mode>=$N_{ph}/M$>>1. Multiple photons($n$>1) interfere, generating wave noise. Shot noise applies when $n$<<1 and the detected photons arrive independently. Power, $P_{sc}$, rather than energy, $\Delta E_{sc}$, is important. The *SNR* is then given by the well-known formula[18,19,20], Eq (10). Multiple, $N_s$, LIDAR scans and $M$ EM modes are averaged as well as integrating the emission over $\tau$ in determining the *SNR*. $P_n$ is the background noise power,

$$SNR_{\delta v_T, N_s} = P_{sc}/(P_{sc} + P_n) \sqrt{MN_s} \sqrt{1 + \delta v_T \cdot \tau} \quad (10)$$

The $\delta v_T$ is the convolved spectral width of $\delta v_p$ with $\delta v_{TS}$. $P_{sc,n}$ define temperatures $T_{sc,n}$ by $P_{sc,n} \equiv k_b T_{sc,n} \Delta f$ for $\Delta f$. $T_{sc,n}$ may be expressed in eV(1eV–11,600°K).

Two *SNR* regimes are delineated by Eq (10): **($P_n$<$P_s$)** The *SNR* does not improve with signal power neither by increasing $E_p$ (a more powerful laser) nor by a higher $n_e$, only by increasing $\tau$. Typically, $\tau$ (=$\tau_{plasma}$) is ~1$ms$ for LP-HP CTS but LIDAR CTS must boxcar average many scans to increase the *SNR* as $\tau$ (=$\tau_{\Delta s}$) is brief. With $N_s$, $SNR_{\delta v, Ns}$ improves to $\sqrt{N_s} \cdot SNR_{\delta v,1}$, implying that $R_p$ must be high, but $R_p$ must also be less than $c/2L_p$ to avoid overlapping emission from consecutive pulses. The maximum rate is $R_p$~75·(2$m$/$L_p$)[MHz] with $N_s$=75,000 scans, for $L_p$=2$m$,



$SNR_{\delta\nu,75{,}000} = 270 \cdot SNR_{\delta\nu,1}$. Choosing a $\tau_{plasma}$, $N_s = R_p \cdot \tau_{plasma}$. Fixing $\delta\nu_p$ to be 2GHz and $\delta\nu_{CTS}$=4GHz then $SNR_{\delta\nu,1}$=2:1, $SNR_{\delta\nu,75{,}000}$= 540:1. Continuous LP-HP CTS *SNR* is higher by ~$\sqrt{N_m}$, with $N_m = L_p/\Delta s \sim 20$. Both systems operate continuously for 1*ms* but CTS SP splits the measurement $N_m$ ways. An $R_p$ of 20MHz is practical, with pulse interval of $\Delta T_p$=50*ns*(15*m*). The single shot $SNR = \sqrt{M}\sqrt{(1+\delta\nu_T \cdot \tau)}$ may suffice for $M \gg 1$. A *post-processed* $\tau_{\Delta s}$ and $\tau_{plasma}$ are both possible. Observing higher BW dynamics with lower *SNR* is available.

$(P_n \gg P_{sc})$ may result from a large ECE background or a low $E_p$. This is dealt with in LP-HP CTS by chopping the source to measure $P_n$ alone for subtraction. A 50% duty cycle halves the *SNR*. LIDAR can accomplish background subtraction, by integrating $P_n$ alone between pulses. In this regime, $SNR \propto P_{sc}/P_n$; raising $P_{sc}$ *significantly* increases *SNR*. LIDAR can leverage this dependence with source powers potentially much higher than gyrotron powers limited by an optical window DTL. Taking the DTL to be 1MW, an $R_p$ of 20MHz has a duty cycle of 1%(=$\tau_p/\Delta T_p$=0.5*ns*/50*ns*) implying the instantaneous $P_o$ can be 100MW with an $<P_o>$ at 1MW, a 100-*fold* improvement in *SNR*! ECE does not contribute to $P_n$ for FRCs, but for tokamaks, overcoming ECE can be paramount. To date, the highest $E_p$ is 14*mJ* in 2*ps*, 7GW of *primal* power. The laser pulse, $E_{NIR}$~1J, so $R_p$ would be low~100Hz. With pulse conditioning, see **SECTION V**, $\delta\nu_p$ can be reduced to 7 GHz lowering $<P_o>$ to 100MW, greatly subduing ECE.

The performance of THz IPS and CTS PS receivers can now be assessed. Schottky detectors/mixers are considered given their wide RF coverage($\mu$wave-THz), high video BW>10 GHz, low noise and high responsivity. The DSB BW $T_{n,DSB}$~1000°K at 300°K and 500°K at 77°K(LiN$_2$ cooled) for $\lambda$~1-2*mm*, approaching the quantum limit, $(hf_o/2)/k_b$=7°K at 1*mm*. Schottky mixers are available to 3THz($\lambda$=0.1*mm*). $P_{n,Het}$ at room temperature is given by Eq (10),

$$P_{n,Het} = k_b T_n \Delta f_{GHz} = 0.0138 \Delta f_{GHz} [nW] \qquad (10)$$

LO and electronic noise contributions, as well as emissive backgrounds, must be included in $P_n$.

Three THz Pulsed Spectropolarimeters are considered:

*Scenario* 1) A CTS Het THz PS with $N_s$=2000. A FM is $E_p(\lambda)$ needed for $P_{sc}=P_n$. Using Eqs (10, 8b), $E_p$(1*mm*) is only 1$\mu$J with $n_e$=4·10$^{19}$*m*$^{-3}$, A$_{2cm}$, $T_H$=10keV, with $\int S_H(\nu)d\nu$=½ and $\delta\nu_T$=4GHz (Eq 6) +$\nu_p$(=2GHz). $P_{sc} \gg P_{n,het}$ with little demand on $E_p$ as shown in **Fig. 2** where an *SNR* of 87:1 is obtained at $E_p$=60$\mu$J, $\lambda_o$=1*mm*.

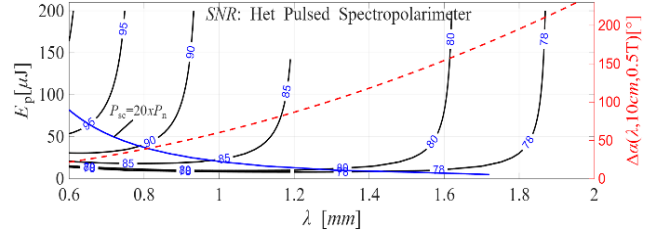

**Fig. 2.** $SNR_{het\ CTS\ PS}(E_p,\lambda)$ for an H plasma, $T_H$=10keV, $n_e$=4·10$^{19}$*m*$^{-3}$, $\delta\nu_T$=6GHz, A$_{2cm}$ and $N_s$=2000. $P_n$=0.08*nW*, $P_{sc}$=4.6*nW* and $SNR$(60*uJ*,1*mm*) =87:1.

Laser specs are modest, $R_p$=200kHz, $E_{NIR}$=3*mJ*, $P_{NIR}$=600W, $\tau_{plasma}$=10*ms*, $\Delta\alpha$(10*cm*) is a robust 40° with $B_{pol}$=0.5T. Improvements would be increasing $P_{NIR}$, $R_p$ and $N_s$ to lower $\tau_{plasma}$.

For $P_{sc} \gg P_n$, the *SNR* depends solely on $N_s$ and $\delta\nu_T$ at a specific $\lambda$. The performance holds until $E_o$ drops below 20$\mu$J allowing a $\tau_{plasma}$=3*ms*. 2kW NIR lasers are available with $R_p$>1MHz making $\tau_{plasma}$=1*ms* practical.

*Scenario* 2) Given the success of het CTS PS, one can ask if a heterodyne technique works for ITS also? Taking $\lambda_o$=0.2*mm*(1.5THz), the Schottky mixer BW is ~60-80GHz. Taking two bands around $\lambda_o$ of 120GHz total to measure the Faraday rotation accurately, the collected emission ranges from 7–35% of $\delta\nu_{ITS}$ over the range $\lambda$=0.1–0.6*mm*, when taken to be that percentage of $\int S_e(\nu)d\nu$. The *SNR* increases by $\sqrt{(122/6)}$=4.5 with the higher $\delta\nu_T$. The $P_{sc}=P_n$ contour corresponds to *SNR*=55:1 and requires $E_p$~0.1*mJ* at $\lambda_o$=0.2*mm* from **Fig. 3**.

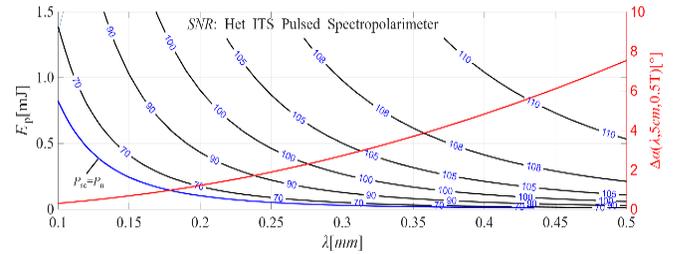

**Fig. 3.** $SNR_{het,ITS,PS}(E_p,\lambda)$, $T_e$=10keV, $n_e$=4·10$^{19}$m$^{-3}$, $\delta\nu_T$=122GHz, A$_{5mm}$=0.2*cm*$^2$, $B_{pol}$=0.5T and $N_s$=200. $P_n$=1.4*nW*, $P_{sc}$=13.7*nW* and $SNR$(1*mJ*,0.2*mm*)=100:1.

The waist radii, $w_{o,ITS}$:$w_{o,CTS}$ are (1:4) scaling as ~$\lambda_{ITS}$:$\lambda_{CTS}$. A pencil beam results, *impossible* to align in a crossbeam diagnostic but for LIDAR, emission is collected regardless of $w_o$ once aligned.

For $R_p$=20kHz, $E_{NIR}$=50*mJ*, $P_{NIR}$=1kW and $\tau_{plasma}$=10*ms*. At $\lambda_o$=0.2*mm*, $\Delta\alpha$=3°/10*cm* at $B_{pol}$=0.5T. Improvements are raising $\lambda_o$ and $P_{NIR}$ to lower $\tau_{plasma}$. Auxiliary measurements are needed for $T_e(s)$ and n$_e(s)$.

For ITS PS, $\tau_p$ can be much less than 0.5*ns*, say 50-100*ps*, without significantly broadening $\delta\nu_{TS}$, simplifying the THz source, see **SECTION V**. The



directional coupler, see **Fig. 7**, that directs the TS emission to the polarimeter, can be nearly 100% efficient instead of 25%, using a dichroic splitter that selectively transmits the pulse but reflects the emission.

*Scenario* 3) A DD ITS THz PS can achieve single shot performance. Schottky diodes have a $P_{n,DD} \sim 10^{-(11-12)}\sqrt{\Delta f}$ [W] for $\lambda_o \sim 0.2\text{-}0.3mm$. For $N_s=1$, the *SNR* is given by Eq (11) and **Fig. 4** plots $SNR_{DD}$ varying $M$, $E$, $A_d$ and $\lambda$.

$$SNR_{DD} = P_{sc}/(P_{sc} + 10^{-11}\sqrt{\delta\nu})\sqrt{M}\sqrt{1+\delta\nu \cdot \tau} \quad (11)$$

An ITS PS measures sightline $B_{\parallel}(s)$ and $n_e(s)$ with a spectropolarimeter for $T_e(s)$ consisting of Schottky diodes for polarimetry, $\mathcal{I}_{p,s}$ filtered with $\delta\lambda_\alpha$ and spectrometer signals $\mathcal{I}_l$ filtered with $\delta\lambda_l$, $l=1,N_d$ for determining $T_e(s)$. Hemisphere concentrators, waveguide antennas and focusing lenses are used to gather the large etendue and spectral widths.

An $E_p=10mJ$ pulse requires $E_{NIR} \sim 0.5J$. A $R_p=1$ kHz is a $P_{NIR}=500W$. Such ultrafast TiSap lasers are available. The $P_{sc}/(P_{sc}+P_n)$ factor is low, this system is noise dominated. Since this is a telescope system, the *SNR* can be increased by raising $A_d$ and/or lowering $f^{\#}$. $SNR_{DD}$ is plotted in **Fig. 4**, $\Delta\alpha=6°/10cm$ at $B_{pol}=0.5T$, $n_e=10^{20}m^{-3}$.

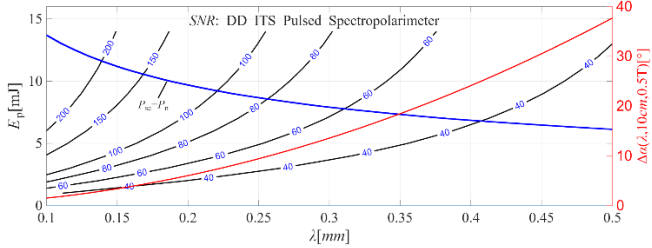

**Fig. 4.** $SNR_{DD,ITS,PS}(E_p,\lambda)$: $T_e=10keV$, $n_e=10^{20}m^{-3}$, BW=$170/\lambda_{mm}$[GHz], $B_{\parallel}=0.5T$, $A_{2cm}=3.1cm^2$, $f^{10}(\Delta\Omega=0.008sr)$. $P_n(0.2mm)=9.2\mu W$, $P_{sc}(10mJ)=9.5\mu W$ and $SNR(10mJ, 0.2mm) = 118:1$.

Background plasma noise is reduced for LIDAR in the optical range, $\sim\sqrt{(P_n\tau_{\Delta s})}$. In the THz range, $P_n$ is not reduced, measurement brevity doesn't matter. ECE is the main noise for tokamaks and THz receivers must deal with this noise in the $P_{sc}/(P_{sc}+P_n)$ term. For FRCs, 2nd harmonic ECE at $2\omega_{ce}$ is typically below $\omega$.

***Summary of instrumental concepts:*** *SNR*s were derived for three THz PS concepts. The het CTS PS, $\lambda>1mm$, uses extant high $R_p$ Yb ultrafast lasers with $P_{NIR}$ to 2kW and $R_p$ to 10MHz, much higher performance than noted. The footprint is small, $\sim 10cm$ dia. mirrors and $T_q(s)$, $n_e(s)$ are obtained but CTS PS can be too sensitive for polarimetry. The het ITS PS, $\lambda>0.3mm$, is faced with much lower spectral brightness but compensations apply. This system must use additional het receivers in addition to the polarimeter to measure $n_e(s)$ and $T_e(s)$. $E_{NIR}$ at 50$m$J and $R_p$=20kHz is a custom 1kW Yb doped laser. The system footprint, <5$cm$ dia. mirrors complements the het CTS PS system as $T_e$, $n_e(s)$ and $B_{pol}(r, z)$ with $(\psi,\chi)(s)$ being modest and measurable for fusion reactor devices. The DD THz ITS PS can make single shot measurements given the large etendue receiver. The footprint is not small, $\sim 40cm$. Ultrashort TiSapphire lasers with $R_p$=1-10kHz make this concept attractive for $T_e$, $n_e$ and $B_{pol}(r,z)$ measurements on a $\tau_{plasma}$=0.1-1$ms$. The large $\delta\nu_{ITS}$ allows $\Delta s$ to be reduced with higher spatial resolution. All three systems are *agile* and can map the 2D MHD equilibria in the poloidal plane.

***LP-HP CTS and CTS SP comparison***: The contrast between a het CTS SP and an LP-HP CTS diagnostic's integration times could not be starker. Nevertheless, by rapid pulsing and averaging over scans the two methods are shown to be comparable. For CTS PS, $\tau_{plasma}$ is *distributed* over $N_m=L_p/\Delta s$ spatial measurements with $\sqrt{N_m}\sqrt{N_s}\sqrt{1+\nu_T \cdot \tau_{\Delta s}} \approx \sqrt{1+\nu_T \cdot \tau_{plasma}}$ for $N_s$=($c/2L_p$)$\tau_{plasma}$ at max $R_p$ in the $(P_n<P_{sc})$ regime. Het CTS PS has the potential to outperform LP-HP CTS by factors of 10 to 100x in the $(P_n>P_{sc})$ regime as $P_o$ can be much greater than $<P_o>$ overcoming a window DTL. The $\Delta f \sim 1/\tau_{plasma}$ for LP-HP CTS while $\Delta f \sim 1/\tau_{\Delta s}$ for het CTS PS no matter how many scans are averaged. Pulsed Spectropolarimetry is an equilibrium *not* a high BW, narrow linewidth diagnostic.

**CTS spectroscopy:** The *distributed* TS emission spectrum, $S_{TS}(\nu,s)$ is the sum of emission from density fluctuations from electrons and ions which *magically* narrows $\delta\nu_{TS}$: $170/\lambda_{mm} \rightarrow 4GHz/\lambda_{mm}$ as $\alpha_s(\pi)$ changes from <<1 to ≥1 with $\int S_{TS}(\nu)d\nu$ reduced by 2. The CTS spectral brightness increases 20x over that of ITS.

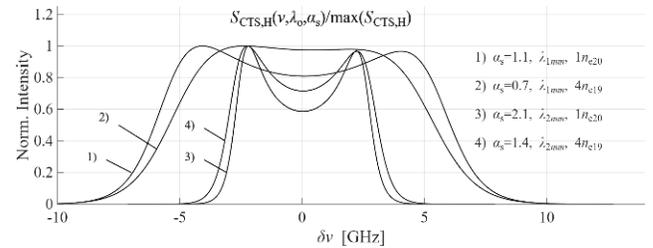

**Fig. 5.** The ion density fluctuation contribution to $S_{TS}$, $S_{CTS}$ as $\alpha_s(\pi) \geq 1$. Four LIDAR scenarios are shown. $T_H=T_e=10keV$. The spectral width, $\delta\nu_{CTS}\sim 1/\lambda_o$. The central dip disappears as $\alpha_s(\pi)$ falls well below 1.

Normalized $S_{CTS}$ are plotted in **Fig. 5** for a 10keV hydrogen plasma, see [8, *p*302]. Peak power scales as $E_p n_e \lambda_o^2/A_d$ by Eq (8b). The emission is structured but the dip will smooth and flatten once convolved with the



pulse's spectrum. For $\delta v_p$=1GHz($\Delta s$=20cm) the spectra shape would be almost unchanged. NBI fast ions and alphas will broaden the spectrum. A vestigial bump on $S_{ITS}$ due to $S_{CTS}$ at $\lambda_o$ disappears as $\alpha_s$ goes to 0.

$S_{ITS}$ profiles for $\alpha_s(\pi) \leq 0.2$ at $\lambda_o \leq 0.3mm$ are shown in **Fig. 6**, generating non-relativistic ITS profiles. An outlier $\lambda_o \leq 1mm$, high $\alpha_s(\pi) \leq 1.5$ with $n_e=10^{20}m^{-3}$ and lower $T_e$=5keV shows a degraded $S_{ITS}$ profile at $\lambda_o$.

For comparison, the ASDEX Upgrade CTS diagnostic[21] is outlined: an O-mode launch, $\theta_{sc}$=122°, $\lambda_o$=2.9$mm$(105GHz), $P_o$=0.5MW, $\tau_{plasma}$=50$ms$, $\Delta s$=6$cm$, $f$=100-110GHz, $N_f$=50, $\Delta f$=0.2GHz, $f_{notch}$=1GHz, peak $\mathcal{J}_{fi}$~15eV, $E_{NBI}$=60-90keV and has successfully measured the effects of NB heating on $T_q$, and energy distributions and slowing down times of NB injected confined fast ions. Instrumental 'solutions' are a 0.5$s$ search to maximize $O_b$, two active and passive CTS receivers to subtract spurious emission. CTS THz PS is not very different except for the high frequency resolution, $\Delta f$=0.2GHz, with $\lambda_o$ = 300GHz, $P_o$=0.5MW and $\tau_{plasma}$=1$ms$, $\Delta s$ = 10$cm$, no notch filter, only one receiver and spatially *distributed* measurements.

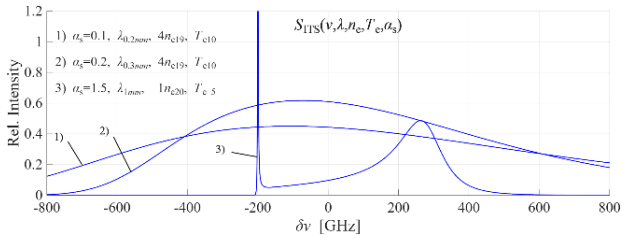

**Fig. 6.** Electron density fluctuation contribution (ITS) to $S_{TS}$ profiles, $S_{ITS}(v)$ is shown. For $\alpha_s(\pi)$<<1, $\lambda_o$~0.2−0.3$mm$. At $\lambda_o$=1$mm$, $\alpha_s(\pi)$=1.5, $S_{ITS}(v)$. is much reduced at $\lambda_o$.

The transition from $S_{ITS}$ to $S_{CTS}$ is due to averaging out the shorter scale ITS fluctuations as $\alpha_s(\pi)$ goes from <0.3 to >1. In the transition (~0.3<$\alpha_s(\pi)$<1) $S_{TS}$ profiles are produced by the interference of ITS and CTS field amplitudes fluctuations. The separation of $S_{CTS}$ from $S_{ITS}$ is never complete but the ratio of $S_{CTS}$:$S_{ITS}$ can be high enough to exclude one over the other for a given *SNR*. Other than density fluctuations, other fluctuating parameters may contribute, electric field, $\delta E$ fluctuations and magnetic field, $\delta B$, fluctuations may significantly alter the $S_{TS}$ profile[22]. CTS PS measures $S_{TS}$ averaged over $\tau_{plasma}$, resolved to $\Delta f$~1/$\tau_{\Delta s}$, if the fluctuations, at the edge say, are stationary over $\tau_{plasma}$, CTS PS is useful.

**V THE THZ PULSED SOURCE**

Significant progress has been made recently in producing powerful THz pulses. OR is used to convert a NIR ultrashort, 50-500$fs$ pulse into a near single cycle THz pulse[23]. The *primal* THz pulse is brief, $\tau_p$<2$ps$, and energetic, $E_p$ ~1$mJ$, *ideal* for LIDAR applications. A record $E_p$=13.8$mJ$ has been achieved using LN crystal[24]. LN has a high nonlinear DFG optical coefficient whereby two NIR EM fields generate a brief DC field. LN also has a high laser DTL and low THz absorption. Phase matching in the conversion process is accomplished using the TPF technique[25]. Conversion efficiencies as high as 3.8% have been reported using cryogenically cooled LN[26]. The range, $\lambda$=0.5-2$mm$(0.15-0.6THz), appropriate for CTS PS is covered. The *primal* spectral width, $\delta v_p$~500GHz, is unusably wide and pulse conditioning is used to reduce $\delta v_p$.

CTS PS requires high $R_p$, afforded by the ultrafast 1.03$\mu$m Yb-based free space or fiber optic laser. Average powers $P_{NIR}$ of 2kW with $E_{NIR}$=20mJ@100kHz to 2mJ@1MHz are stock with higher power a matter of resources. Recently[27] OR with LN was demonstrated with a 10W Yb laser, $R_p$=25kHz, $\eta$=1%.

For ITS PS, $\delta v_p$ can be >>2GHz given $\delta v_p$<<$\delta v_{ITS}$, $\Delta s$ is determined by $\tau$ alone, Eq (4). CTS PS requires $\delta v_p$~2GHz. Reducing $\delta v_p$ by bandpass filtering would incur an unacceptable loss. The large $\delta v_p$, however, works to advantage if THz waveform synthesis is used. In its simplest form, waveform synthesis narrows $\delta v_p$ by splitting the pulse into a *uniform* train of $N_{THz}$ THz pulses, $dE_p$=$E_p$/$N_{THz}$ with delay, $\delta$. Then $\delta v_p$ is reduced to 1/($N_{THz}\delta$) with $f_o$=1/$\delta$ and $\lambda_o$=$c\delta$. A narrow filter would block nearly 100% of the *primal* pulse but replicating the *primal* pulse $N_{THz}$ times allows a pulse train to pass the filter almost unimpeded. Any wavelength of significant amplitude in the spectral envelope of the *primal* THz pulse, spanning three octaves, 0.5−2$mm$, can be developed into an energetic, spectrally narrowed THz pulse. Higher harmonics may be included with the fundamental, BP filtering may be needed.

Pulse train generation is typically done at the NIR wavelength (800 or 1030$nm$) using a sequence of crystal retarders, each producing a specific delay and doubling $N_{NIR}$[28]. With $m$ in-line(on-axis) delays that progressively double in thickness, a train of $2^m$ pulses is produced with *uniform* delay, $\delta$, set by the thinnest crystal[29]. There is an inefficiency, the $2^m$ sequence consists of two orthogonally polarized $2^{m-1}$ sequences, splitting $E_{NIR}$ in half. The pulse duration, $\tau_p$ is $2^{m-1}\delta$.

Many LIDAR opportunities are afforded by an ultrashort *primal* pulse. A $\Delta s$=2$cm$ or less can be realized if $\tau_p$=100$ps$, with $P_o$=4MW of a 400$\mu$J pulse. The larger $\delta v_p$=10GHz may be acceptable for ITS PS. For



Δ$s$~10$cm$, $τ_p$ can range from 300-700$ps$ and maintain a Δ$s$~10$cm$, Eq (4) with an improved $SNR$ and does not unduly widen $δν_T$. $P_o$ can be increased significantly at the expense of a wider $δν_p$, although a higher $P_o$ is best achieved with a higher $E_p$.

An outstanding issue in achieving the highest rep rates is the thermal management of the LN crystal heat load at high average pump powers due to THz absorption. The heat loading can be significantly reduced by cooling with the added benefit of increasing $η$. $P_{NIR}$ approaching 1kW@100kHz pulse rates have been achieved with customized LN crystal geometry to optimize heat management[30]. Broadband AR coatings or structures lowering Fresnel losses are beneficial.

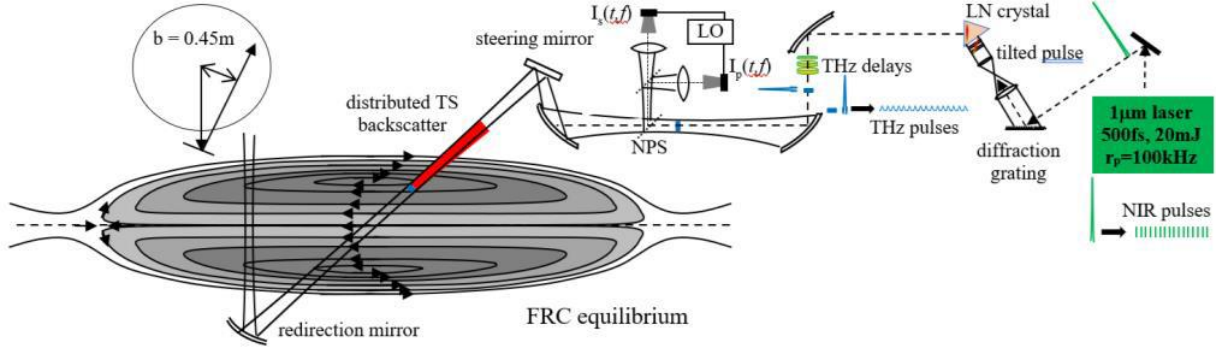

**Fig. 7.** A THz heterodyne Pulsed Spectropolarimeter diagnostic applied to an FRC equilibrium is illustrated. A spectrally narrow pulse is generated by OR and waveform synthesis. A non-polarizing beam splitter (NPS) separates the outgoing pulse train from the incoming TS backscatter which is diverted into a simple polarimeter generating the $I_{s,p}(t,f)$ signals from mixing the emission with a local oscillator. A plane steering mirror illustrates the aiming of the sightline into the plasma. The sightline can be redirected by an in-vessel mirror to measure fast ion distributions at a specific impact parameter, $b$.

*Summary:* The THz source for a Pulsed Polarimeter needs a wide ~GHz linewidth. Transient intense THz pulses generated by OR using LN are ideal given THz waveform synthesis as a means to narrow the spectrum. Present technology can provide THz pulse trains which meet the bandwidth, mode quality and polarization requirements. DD single pulse PS with $E_p$ of several $mJ$s and rapid scan heterodyne PS with $E_p$<0.1$mJ$ are both technologically viable. At present, extant ultrafast NIR lasers with 2kW average power can provide high $SNR$ measurements at kHz rates with spatial resolutions of 10$cm$ in both versions of PS.

## VI AN FRC THZ PS AND ITS PERFORMANCE

The FRC plasma shown in **Fig. 7**, together with external coils and vacuum chamber has several technologically appealing aspects[6]: 1) *Geometric simplicity*: no central column or solenoid, the FRC magnetic field topology is simply connected; 2) *Magnetic simplicity*, little or no toroidal field and no toroidal field coils; 3) *Self-organized loop voltage* that has been sustained *indefinitely* with NBI and end biasing; 4) Linear chamber, FRC has *freedom to be translated*; 5) Separatrix separating closed field lines from open field lines with open field lines (scrape-off layer) forming a *natural divertor*; 6) *Magnetic trap* using simple end coils and 7) a stable *high* <β> equilibrium. The confinement time must be improved for FRCs to be taken seriously.

A THz heterodyne pulsed spectropolarimeter implemented on an FRC equilibrium is shown in **Fig. 7**. Only single port access is required. The highly prolate shape of the FRC allows a determination of the midplane $n_e(r)$ and $B_z(r)$ profiles without using radial sightline. The tilted sightline is symmetric about the midplane, passing through ($r=0,z=0$) at an angle $γ$ to the machine's axis. The measurement set is shown in **Fig. 8(a-d)**. The Faraday rotation: $α_{tot}(s)=2C_{FR}·λ_o^2∫n_e(s)(B_z(s)·\cos γ + B_r(s)·\sin γ)·ds$, $γ=∡(\hat{s},\hat{y})$. $B_r(s)$ is neglectable. $B_∥(s) ≈ B_z(r(s),0)\cos(γ)$ with $r=(s-s_o)\sin(γ)$, $s_o=s(r=0)$. The measured $B_∥(s)/\cos γ$ faithfully represents $B_z(r)$. The perpendicular field along the sightline is $B_⊥(s)≈B_z(r(s),0)\sin(γ)$, neglecting $B_r(s)$.

Another magneto-optic activity given by cold dispersion theory is the CM effect, a retardance producing an ellipticity $Δχ=½(N_X–N_O)k_o Δs$ for the {O,X} eigenmodes when $\mathbf{B}·\hat{s}=0$, given by Eq (15),

$$Δχ ≈ ω_{pe}^2 ω_{ce}^2/2ω^4 · kΔs · \mathbf{B}_{perp}^2/B^2 \sin(2β_E) \quad (15)$$

where $β_E=∡\mathbf{E},\mathbf{B}_{perp}$, $Δχ=0$ if $\mathbf{E}$ is ∥ or ⊥ to $\mathbf{B}_{perp}$. For FRCs, $\mathbf{B}_{perp}$ is $\mathbf{B}_⊥$. For LIDAR the line integrated $Δχ$ is doubled, Eq (16a) at max effect, $β_E=±π/4$. $|B_⊥|(s)$ Eq (16b), by inverting (16a).



$$\chi_{\text{tot}} = (2) \cdot C_{\text{CM}} \lambda^3 \int_o^s ds'\, n_e(s') B_{perp}^2(s') \quad (16a)$$

$$B_{perp}^2(s) = 2 \cdot 10^{10}/\lambda^3 n_e(s) |\Delta\chi_{\text{tot}}(s)/\Delta s| [T^2] \quad (16b)$$

As long as the ellipticity $|\chi_{\text{tot}}| \ll 45°$, $\alpha_{\text{tot}}(s)$ is not affected, see **APPENDIX B**. This allows the use of a less demanding 2 channel polarimeter.

The QO beam can be extended, with little loss, using an in-vessel focusing mirror. The second throw, as shown in **Fig. 7**, is insensitive to $B_{\text{pol}}$ but can be tilted out of plane, to measure the density distribution of confined fast ions, $n_{\text{FI}}$, vs impact parameter, $b$. Any azimuthal drift velocity, $v_\phi$ of the thermal ions appears as a 'small' spectral shift, $\Delta f_\nu(=v_\phi/\lambda_o)$ on $S_q(f,s)$. The fast ions can be isolated in $S_{\text{CTS}}(\nu,s)$ and a fast ion slowing down time determined by modulating the NBI ON/OFF.

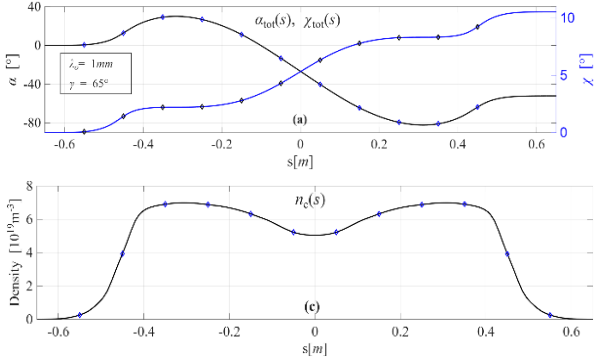
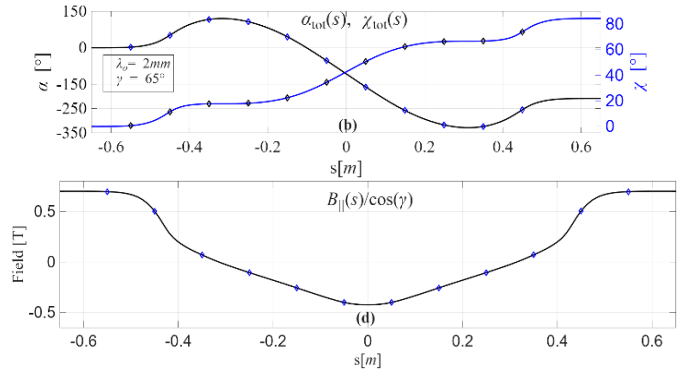

**Fig. 8** Simulated polarimetry measurements, $\Delta s$=10cm: **(a,b)** line integrated SOP($\alpha_{\text{tot}},\chi_{\text{tot}})(s)$, $\gamma$=65°, $\lambda_o$=[1,2]mm, the CM effect is strong, further analysis is required(**Appendix B**), **(c,d)** sightline $n_e(s)$ and $B_\parallel(s)/cos(\gamma)=B_z(r(s))$ profile.

A PP measurement set is shown in **Figs 8a-d** for an FRC simulation. The highly prolate FRC shape, **Fig. 7**, allows a determination of the mid-plane $n_e(r)$ and $B_z(r)$ profiles using sightlines tilted to the machine axis by $\gamma$, in addition, to *distributed* 1D $S_q(\nu,s,\hat{s})$.

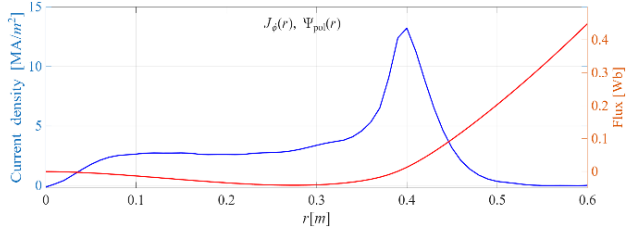

**Fig. 9.** Mid-plane azimuthal current density, $J_\varphi(r)$ and poloidal flux function $\Psi_{\text{pol}}(r)$.

Once $B_z(r)$ is fit with a differentiable curve, the azimuthal current density $\mu_o J_\varphi(r)=\partial B_z(r)/\partial r$ and poloidal flux function, $\Psi(r) = \int_0^r 2\pi r' dr' B_z(r')$ can be obtained, see **Fig. 9**.

Refractive effects must be considered, see **Fig. 10**. Ray tracing indicates that an initial trajectory at $\gamma$ produces an axial offset of ~cms from (0,0) with a mostly straight trajectory at a shallower angle. The launch can be displaced axially and angle adjusted to center the beam on in-vessel mirrors. Longer wavelengths become useable at larger $\gamma$, with $\alpha_{\text{tot}} \sim \lambda_o^2 cos\gamma$ and $\chi_{\text{tot}} \sim \lambda_o^3 sin\gamma$. **Fig. 10** shows a $\gamma$=80° trajectory at $\lambda_o$=2mm closely following the unrefracted trajectory. An outlier beam launched at $\gamma$=45° with $\lambda_o$=2.85mm illustrates refraction so strong that the beam is bent horizontal, penetrating only 20cm, nevertheless, CTS emission will be gathered.

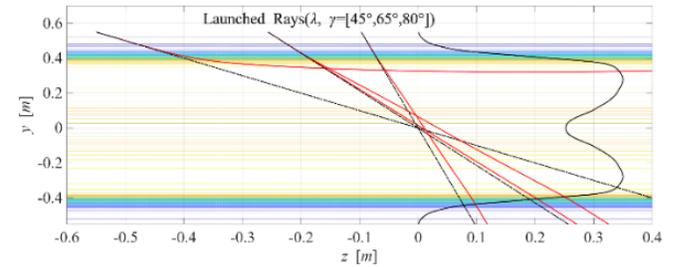

**Fig. 10.** Four rays are traced with geometric ray tracing using $N_O$: $\gamma(\lambda_o)=\{45°\{2.85mm\}, 65°\{1,2mm\}, 80°\{2mm\}\}$. $n_e(r)$ is plotted at half value as a visual aid. Un-refracted rays are shown(dotted).

Double refraction occurs to some degree. The pulse is a superposition of $\hat{e}_{+/-}$ polarized pulses which separate due to $(N_+ - N_-) \neq 0$ but this is negligibly small and, in any case, reciprocal.

***End-on trajectories*** are just as efficacious for exploring FRC equilibria. The FR effect dominates on near axial sightlines allowing a straightforward interpretation of 2D scans. Little is known about the X point and field line curvature of the equilibrium, also true of tokamak X points. A wavelength and THz PS dedicated to intensive mapping of the X point region can be chosen.



The THz pulse train in **Fig. 7** is designed to have $2^{m-1}$ pulses with a $\tau_p=2^{m-1}\delta$~$0.5ns$, narrowing $\delta\nu_p$ to 2GHz. Possible birefringent crystals for optical delays are given in TABLE I. Other relevant properties are DTL, dispersion, absorption, availability, *etc*. Crystal diameters are $25mm$ or larger.

TABLE I

| Material | $N_o$, $N_e$ ($\lambda=800nm$) |
|---|---|
| Calcite | 1.6487, 1.4822 |
| $\alpha$-BBO | 1.6579, 1.5379 |
| YVO4 | 1.9721, 2.1856 |

Calcite's differential birefringence is high, $(N_o-N_e)/c$ =$5.55ps/cm$ @$0.8\mu m$. A $\Delta=0.151cm$ thick crystal produce two orthogonally polarized pulses with a delay of $0.84ps$ ($f_o=1.2THz, \lambda_o=250\mu m$). Ten crystal delays with thicknesses, [$2^0\Delta$, $2^1\Delta$, $4\Delta$, $8\Delta$,…,9.69,19.4,38.7, $2^9\Delta$ (=$77.3cm$)] produce two orthogonally polarized sequences of $N_{THz}=512(=2^9)$ with duration $\tau_p=0.43ns$. Starting the sequence at $4\Delta$ or $8\Delta$ by removing the first 2 or 3 delays will generate ($f_o=300GHz, \lambda_o=1mm$) or ($f_o=150GHz, \lambda_o=2mm$) with $\tau_p$ unchanged.

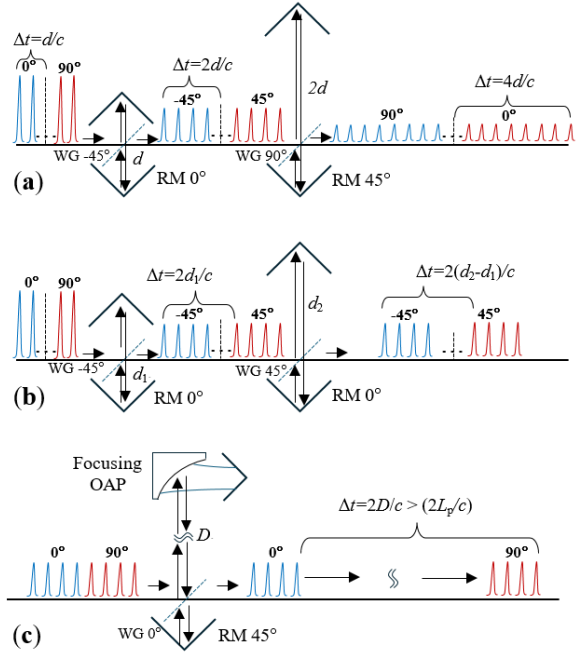

**Fig. 11** Illustrates counterpart QO delays to birefringent crystal delays. **(a)** One/two MP–*like* delays, of $\Delta t=2d/c$ and $4d/c$, replace one/two xtal delays doubling or quadrupling $N_{THz}$, **(b)** an arbitrary delay, $\Delta t=2(d_2-d_1)/c$ is produced with two MP-like delays, **(c)** an arbitrarily long delay widely separates two sequential scans by $\Delta t=2D/c$ (>$2L_p/c$).

This crystal delay scheme is limited as the crystals become impractically long, $9.69cm$ is probably the limit although crystals can be physically joined together in pairs. Absorption may begin to play a role as well. The $8^{th}$-$10^{th}$ delays can be implemented as free space optical or QO delays in the THz range. **Fig. 7** shows both NIR and THz delays in the FRC SP diagnostic layout.

Longer delays can be free space Michelson or polarization-based MP-*like* delays with 100% throughput, using QO components: WG polarizers, RMs and focusing lenses or mirrors where needed as shown in **Fig. 11**. **Fig 11a** replaces the $9^{th}$ and $10^{th}$ birefringent delays with $d=3.2cm$. The $8^{th}$ birefringent delay, $d=1.6cm$, is too short for a *practical* QO delay. **Fig 11b** uses two *practical* QO delays for one arbitrary delay. **Fig 11c** widely separates the two orthogonally polarized pulse trains, $\Delta t>L_p/c$, so as to avoid overlapping emission. Optical sources can precisely align QO delays.

QO beams are guided free space Gaussian EM modes, of arbitrarily polarization. The beam spreads radially with $s$ due to diffraction and optics: lenses and mirrors, are used to *re*-focus the beam to a waist, of radius $w_o=w(s=0)$ where spreading again takes over. The QO components are typically $6-15cm$ in diameter with tolerable truncation losses. Manufacturing tolerances for mirrors/lenses are modest. The performance of WGs and mirrors is near ideal. For inline systems (cylindrically symmetric) polarization control is robust, corners introduce polarization errors. A $2w_o$ beam width keeps truncation losses around 0.3% *per component*.

The radius of curvature of the phase front for a Gaussian mode is $R(s)=s(1+S_R^2/s^2)$, where $S_R=\pi w_o^2/\lambda$ is the Rayleigh length. For $s\in[-S_R,S_R]$ the beam is well approximated by a plane wave. The beam radius $w(s)=w_o(1+s^2/S_R^2)$ converges to and diverges from $w_o$ asymptotically with angle $\theta_g=\lambda/\pi w_o$.

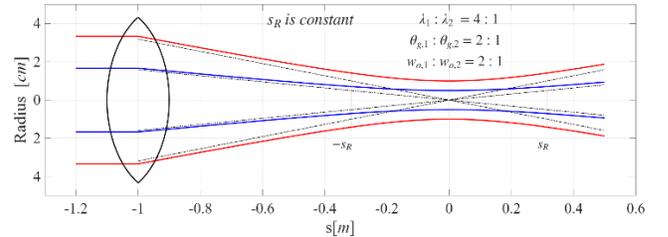

**Fig. 12.** Two collimated beams with aperture ratio 2:1 and $\lambda_1:\lambda_2=4:1$ are focused by a QO lens to the same waist position at $s=0$. $S_{R,i}$ are identical. $\lambda_1=1mm$ and $\lambda_2=0.25mm$.

A two GB system is shown in **Fig. 12**. If $w_{o,i}^2/\lambda_i$ is constant, the two beams can be guided in one QO system.

Multi-$\lambda$ pulse trains can be generated as in **Fig. 13** with $\lambda_1:\lambda_2 = 4:1$ and beam radius reduced by 2 for the $\lambda_2$ beam in the delay as required in **Fig. 12**. This suggests that ITS and CTS het Pulsed Spectropolarimeters can be



combined into one instrument with both $T_e(s)$ and $T_q(s)$ measured simultaneously on the same sightline.

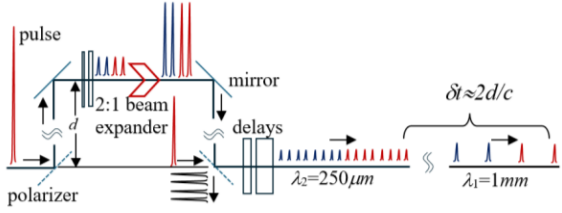

**Fig. 13.** One pulse to two pulse trains with $\lambda_1:\lambda_2=4:1$. A polarizing optical delay contains the first two birefringent delays and a beam expander (intensity is illustrated). Pulse trains are delayed by $2d/c$. A loss of 25% results.

***The polarimeter:*** $E_p$ is evenly split between two orthogonally polarized pulse trains. This can be exploited by delaying the lagging pulse train by $\tau_D \sim 10$'s $ns$ ($>2L_p/c$) and reintroducing the beam into the beamline using polarizing beam splitters producing two *equivalent* scans and using all of $E_p$. The redundant measurement from this *dithered* source improves *SNR* by $\sqrt{2}$ and provides an instrumental check of the polarimeter.

A two detector polarimeter assumes detectors are 'balanced' in responsivity and dark current with an insensitivity to mechanical drift. A single detector polarimeter is automatically balanced. LP states at $\pm 45°$ to the polarizer's axis produce a *direct* measurement of $n_e B_\parallel(s)$ by a *difference-over-sum*, Eq (17) for $\alpha \ll 45°$,

$$n_e B_\parallel(s) \propto (I_{45} - I_{-45}(\tau_D))/(I_{45} + I_{-45}(\tau_D)) \quad (17)$$

where the measurements are separated by $\tau_D$. The *non*-linear formula is used for larger $\alpha_{tot}(s)$. Two such outputs would use all of $E_p$, increasing the *SNR* by $\sqrt{2}$.

For high rep rates of 100kHz to 10 MHz, the Yb doped fiber optic lasers at $1030\mu$m have $P_{NIR}$ to 2kW with $E_{NIR}$ of $20mJ$@100kHz.

NIR pulse trains can also be generated using the Vernier effect regenerative amplifier technique[31], producing a train of predetermined $N_{NIR}$ pulses and delay, $\delta$ before the compressor thereby making $\tau_p$ and $\lambda_o$ programmable. At present, $R_p < 100$kHz and $P_{NIR} <$2kW.

The directional coupler in **Fig. 7**, a 50:50 NPS, introduces a 75% loss. Other types of directional couplers that *co-align* the pulse sightline to the collection optic axis might be based on spectral selectivity, passing a narrow pulse spectrum but reflecting the wider emission spectrum, generating a spectral hole much the same as LP-HP CTS. A promising approach uses a directional coupler that can be transiently switched between transmissive to reflective (a transient mirror) fast enough to allow the pulse to pass but the emission to be nearly 100% reflected[32].

A $\delta v_p$ of 2 GHz can significantly broadening $\delta v_T$ which also increases the *SNR*. The coarse spectral resolution, $\Delta f$=2GHz, is appropriate for confined fast ions with an extended spectrum well beyond $\delta v_{CTS}$ but at lower $n_{FI} \ll n_e$. For larger devices, $\delta v_p$ can be reduced with $\Delta s$ increased, keeping $\Delta s/L_p$ constant. Give a $\tau_{\Delta s}$, the frequency resolution is $\Delta f=1/\tau_{\Delta s}$, but according to (4), there is scope to adjust $\tau_p$ and $\tau$ for a fixed $\Delta s$ with the number of frequency bins, $N_f=\delta v_T/\Delta f$. Decreasing $\tau_p$ and increasing $\tau$ for fixed $\Delta s$, increases $N_f$ but doesn't improve results as $\tau_p$ broadens the resolution. Present day DAQs have BWs of 10GHz allowing digitization and *real-time* analysis of the *distributed* emission spectra without filter banks. $I_{p,s}(s_i)$ $S(v,s_i)_l\Delta f$ for $l=1,N_f$ can be resolved using digital Fourier transforms.

## VII H<small>ET</small> PP <small>AND</small> P<small>OLARIMETRY</small> C<small>OMPARISON</small>

Polarization is one of light's most versatile degrees of freedom. To be able to probe an inhomogeneous anisotropic magnetized plasma in arbitrary directions *and*, through TS, remotely measure a *local* response to a polarized light source is very powerful. THz pulsed sources with initial SOP($\psi_o,\chi_o$) oriented at $\hat{s}$ can bring forth the *rich tapestry* of the MHD equilibrium structure.

CW polarimetry, one of a few field sensitive *advanced* diagnostics[33] provides an instructional comparison to a het PP (a pulsed laser). PoPola polarimeters with poloidal plane sightlines are proposed to ascertain $\mathbf{B}_{pol}(r,z)$. Polarimetry, as simple as diagnostics come, has an Achilles' heel: requiring sightline termination, either on an optical port or a retroreflector embedded in a boundary wall. Retroreflectors are a source of polarization distortions due to refractive misalignment, surface degradation and optical aberrations[34]. Terminated sightlines are fixed in space and have alignment issues typical of crossbeam diagnostics. The spatially constrained measurement is countered using tomography: deploying many sightlines of order 10. A pulsed polarimeter does not require sightline termination, having directional freedom. The diagnostic footprint of a polarimeter is small, as is the case for a heterodyne PP. A 20$cm$ dia. mirror is felt sufficient given $\lambda_o$=1$mm$ on an ITER sized device. The most compelling advantage of het PP over CW polarimetry are the *distributed* measurements of $B_\parallel$ and $n_e$ compared to a chord averaged product $<n_eB_\parallel>_{Lp}$. Usually, interferometry(the "simplest" *advanced* optical diagnostic) is paired with polarimetry to provide sightline $<n_e>_{Lp}$ but such combined systems have been rejected for ITER. One pulsed polarimeter excels over



multiple paired polarimeter/interferometers. With sightline steering, nearly continuous coverage can be achieved. Viewing het PP as a pulsed laser is apt as a sightline can be extended almost indefinitely by refocusing and throwing the beam in a new direction provided centering on the redirecting mirror is possible.

CW polarimetry must balance a larger $\lambda$ yielding a larger Faraday rotation against beam refraction. Line integrated FR angles are kept small to keep beam displacements at the retroreflector/detector tolerable. This is not the case for PP, large rotations are desired and refraction or close approach to a cutoff is not an issue.

Polarimeters can also suffer a retardance from the presence of a strong $B_{\text{perp}}$, perhaps the strong $B_\varphi$ in a tokamak. The CM effect produces a sightline integrated ellipticity, $<\chi>_{Lp}$, along with $<\alpha>_{Lp}$. PoPola measurements on ITER at $\lambda=118\mu m$ are detailed in [34]. The *distributed* SOP ($\alpha$, $\chi$, $I$)($s$) can be measured with a het PP implementing a time division 4 channel Stokes' vector polarimeter (see **APPENDIX B**). When both effects are comparable, $\alpha(s)$ may become entangled with $\chi(s)$[35] as the parameter space of ($\psi,\chi$), the Poincaré sphere, is compact. The mapping of ($\alpha,\chi$) to ($B_\parallel$, $B_\perp$) is no longer trivial. CW Polarimetry must contend with chord averaged ($<\alpha>_{Lp},<\chi>_{Lp}$) not necessarily $\propto(<n_eB_\parallel>_{Lp},<n_eB_{\text{perp}}^2>_{Lp})$. Since retardance can appear as a pure Faraday rotation, disentangling the two magneto-optic activities is not possible unless the sightline history $(\alpha,\chi)(s)$ is known, then $(B_\parallel, B_{\text{perp}})(s)$ can be determined. Field line curvature also generates a sightline $B_{\text{perp}}$. CM is usually small due to the $\lambda^3$ factor, not the case for long wavelength CTS PP. CTS PP as a PS is primarily useful for only *local* ion spectra, on a tokamak being too sensitive field sensing. ITS PP can be useful for 2D mapping of $B_{\text{pol}}(r, z)$ with $\lambda_o \sim 0.1$–$0.3 mm$.

Another handicap for polarimetry with plasmas attaining thermonuclear parameters ($T_e >> 1$ keV) is the need for *warm* (finite–$T_e$) dispersion theory Warm CM and Faraday effects[36] require the *local* $T_e$ for their interpretation. Sightline $T_e(s)$ can be provided by het CTS PS with $T_e/T_q(s)$ ratio assumptions or using ITS PS. For $T_e$ 10keV, the depolarization term in Lienard Wiechert potential[37] in a LIDAR geometry becomes apparent with a 5(12)% depolarization effect at 10(20)keV. In addition to this, the superposition of many random scattering events over a large $\delta\nu$ naturally depolarizes the TS emission. This effect peaks at $\theta_{\text{sc}}=90°$ and is so evident that proposals[36] to measure $T_e>10$keV based solely on intensity have been advanced.

For LIDAR, $\theta_{\text{sc}}=180°$ the effect is not strong. LIDAR TS is a robust *advanced* diagnostic technique.

An arbitrary initial SOP($\psi_o,\chi_o$) is a *key* feature of THz PP. Source dithering has been mentioned but more general schemes are possible. The SOP history can be represented on the Poincaré sphere, see **Fig. B5.**

Diagnostic completeness ("*diagnostic synergism*") encapsulates the PP/PS techniques. *Local* measurements of $B_{\parallel,\text{perp}}(s)$ can be corrected if $T_e(s)$ is also measured. The $\alpha_{\text{tot}}$ and $\chi_{\text{tot}}$ may be individually suspect when the FR/CM effects are strong, but the sightline history (*spatial continuity*) allow the effects to be uncoupled or '*disentangled*'. *Local distributive* measurements give continuity in space and the rapidity of scanning($\tau_{\text{plasma}}$), gives continuity in time. Beam steering enlarges the spatial continuity to most of the poloidal plane, see **Fig. 14**. The simultaneous and coincident measurements of $n_e(s,t)$, $n_{q,\text{FI}}(s,t)$, $B_{\parallel,\text{perp}}(s,t,T_e(s))$, $T_e(s,t)$, $T_q(s,t)$, $T_{\text{FI}}(s,t)$ are mutually synergistic and complete. Larger devices, fusion reactors, can substantially improve the measurement capability. All of these plasma parameters are *key* inputs for Grad-Shafranov equilibrium solvers, $P_q$, $P_e$, $J$ at ($r,z$) and $n_{q,e}(r,z)$ for transport codes.

## VIII FUSION DEVICES AND DIAGNOSTIC AGILITY

Fusion reactor capable diagnostics are *advanced* diagnostics. These diagnostics provide *real-time* feedback for maintaining stability and control of the burning plasma and optimizing performance.

Advanced optical diagnostics should use the native ion species, those based on atomic emission from particle beams suffer a fundamental limitation in signal strength which fades with depth into the plasma and often competes with edge emission. The diagnostic footprint should be small, crossbeam diagnostics require several ports, THz Pulsed Spectropolarimeters, just one port. Wavelengths in the $\mu$wave–THz range mitigate first mirror damage from direct proximity to the plasma. Background emission, $P_n$, is low in the THz range, cutoffs are more easily avoided. THz PS as an optical diagnostic is in the company of Thomson scattering, polarimetry, reflectometry and passive ECE. All must mitigate refraction, severely limiting their range and usefulness.

LIDAR TS techniques provide *remote, non-perturbative*, *snap-shots* of *distributed* spectra $S_{q,e}(\nu)(s)$, SOP($\psi,\chi$)($s$), intensity $I(s)$ directly related to *key* plasma parameters $n$, $B_{\text{pol}}$, $T_e$ and $T_q$ using *one* detection system. The brevity of LIDAR (6.6*ns/m*) either reduces or dominates background noise. Measurements are suitable



for *real-time* feedback with scan rates of 1kHz, 10*cm* spatial resolutions and high *SNR*.

*Distributed* $B_\parallel(s,T_p)$ for a single scan or running average time, $T_p$ is given by (18) where $\alpha^*(s)$ is a smooth differentiable curve fit to $\alpha_{tot}(s_i)$,

$$B_\parallel(s_i,T_p) = C_B/n_e(s_i,T_p) \cdot d\alpha^*/ds(s_i) \quad (18)$$

and $n_e$ is given by (19), $C_B$, $C_n$ are constants,

$$n_e(s_i,T_p) = C_n \sum I_{s,p}(s_i,T_p) \text{ or } C_n \sum I_{45,-45}(s_i,T_p) \quad (19)$$

The reduction of measurements to useful parameters in *real-time* is possible since the algorithms are direct. Non-linear formulas or fits to the spectral intensities, $\mathcal{I}_{\lambda_l}(s_i), l=1,N_\lambda$, can use look up tables to determine $T_q(s_i)$ in *real-time*.

CTS is performed using LP-HP gyrotrons with infrastructure and manpower costs so high that CTS diagnostics try to use existing ECRH sources where available and just add a receiver. ECRH sources are tuned to high ECA which is not what one wants for CTS!

CTS is presently the only diagnostic technique able to measure the *local* density and energy distribution of confined *alphas* produced by fusion reactions. A low *alpha* slowing down time to confinement time ratio is crucial for the success of a burning plasma. A high anomalous loss of fast ions would significantly affect the performance of a fusion reactor. By keeping the central spectrum, *local* D,T:He(*s*) ratios can be measured.

LIDAR sightlines are only constrained by the port's FOV. With a steering mirror, the diagnostic is *aim-and-shoot* with one receiver for all sightline measurements. Beam refraction is tolerated and with a judicious optical design the beam can be extended using in-vessel mirrors as $\Xi(=\lambda_o^2)$ is constant.

This work introduces a new diagnostic concept, that of *diagnostic agility*. One might ask of a diagnostic, given the long discharge durations of ITER and future fusion reactors that a diagnostic providing *real-time* temporal feedback might also be maneuverable to provide spatial feedback as well. For optical diagnostics, this might amount to the movement of a mirror on a *ms* time scale. *Agility*, effectively replicates a diagnostic many times over, saving a huge effort and cost in providing more diagnostic coverage. Diagnostics with crossbeam geometries as well as fixed sightline diagnostics cannot be *agile*.

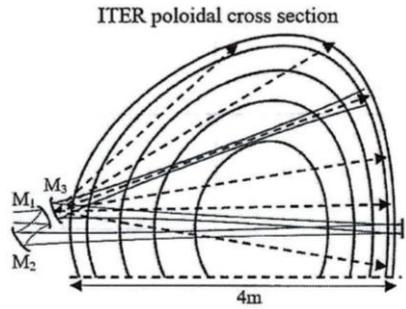

**Fig. 14.** ITER is used to illustrate a folded system using a reflective flat on the center stack. For $\lambda \sim 1mm$, $\sim 20cm$ dia. mirrors suffice. Refraction is small from $M_2$ to $M_3$ with radial profiles of $n_e$, $T_q$ sensed. The initial LP state and $\lambda$ can be chosen to avoid and be above EC absorption. Unfortunately, the CM effect is overwhelming and only $S_q(\nu)$ is measured. After $M_3$, refraction doesn't matter. A scan time of 80*ns* limits $R_p$ to 10MHz. The *SNR* can be high~200:1, with $N_m$=120, $\tau_{plasma}$=1*ms*. In 1*s*, many sightlines can be sampled by rotating $M_3$. A 10kHz BW would be fast enough to sense and track MHD instabilities such as NTMs with reduced *SNR*. Het IPS PP with $\lambda \sim 0.25mm$ is appropriate for 2D mapping of $B_{pol}(r,z)$, $n_e(r,z)$ and $T_e(r,z)$.

*Agile* techniques can spatially resolve control inputs, such as edge fueling or auxiliary heating deposition profiles as well as localizing instabilities. The sightline steering *on demand* extends the diagnostic coverage from the plasma edge to the peak $n$ and $T_q$ at the m.a. with sightline dwell time set by a desired spatial resolution and measurement accuracy. 2D maps of poloidal parameters are possible with 20 sightlines at 10Hz rates with 10*cm* spatial resolution and *SNR*~100:1.

**Fig. 14** illustrates the *agility* concept on the ITER device. The $M_3$ mirror steering action is improved by a double pass system to the central column and back. Not only does this improve the poloidal plane coverage but also provides peak plasma profiles through the m.a. with every scan in addition to the sweep. Continuous scanning An $R_p$ of 10 MHz is technologically feasible.

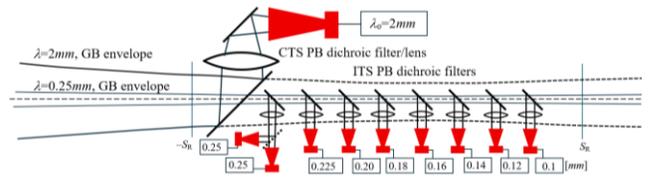

**Fig. 15.** Dual $\lambda$ {0.25,2}*mm* Gaussian beams are shown with filtering components within the Rayleigh range. The two $\lambda$s are 8:1 ratio, The filters consist of 1 large dichroic filter to separate the lower CTS spectrum from the ITS spectrum and 8 dichroic filters to dissect the ITS spectrum into 1 polarimetry and 7 intensity channels.

Using custom laser sources with sufficient $P_{NIR}$, the pulse train can be split into two wavelengths and with



a partitioned multi-LO receiver, **Fig. 15**, measuring both $T_e(s)$ from ITS and $T_q(s)$ from CTS spectra.

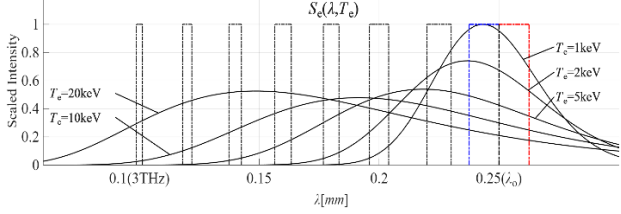

**Fig. 16**. The range of Schottky mixers ends at 3THz(0.1*mm*). Choosing a $\lambda_o$=0.25*mm* to 0.3*mm* allows the LOs to better bracket the emission profiles for $T_e$ up to 20keV. Two 60 GHz filters(120 GHz wide) are used for polarimetry to either side of $\lambda_o$. LO filter widths become wider at shorter wavelengths.

ITS emission spectra at relativistic temperatures[38] are shown in **Fig. 16**. A dual wavelength 10 LO het QO receiver, **Fig. 15**, resolves $S_e(T_e)(s)$ up to 20keV. More BP filters–LOs modules can be added as needed.

## IX CONCLUSION

The Pulsed Spectropolarimetry technique has been detailed and illustrated on an FRC equilibrium. Only modest THz pulsed source development is needed beyond the present state for generating THz pulses with the required wavelength, pulse energy and spectral bandwidth. Extant ultrafast NIR pulsed lasers of 1kW class suffice with demonstrated THz pulses generation by optical rectification on crystalline lithium niobate. THz waveform synthesis determines the central wavelength and spectral width.

The wavelength range is appropriate for remote sensing using CTS and ITS backscatter. *Key distributed* plasma parameters are sensed. $B_\parallel(s)$ and $B_\perp(s)$ result from differencing line-integrated LIDAR measurements by means of sightline optical dispersion. *Local* $n_e(s)$, $T_e(s)$ and $T_q(s)$ are obtained directly as on the JET TS LIDAR diagnostic. TS backscatter is a robust, high $T_e$, diagnostic agent that encodes an incredible amount of spectral information by varying $\lambda_o$ and location $(r(s),z(s))$ from plasma edge to m.a. and *key* parameters through polarization$(\psi,\chi,I)(s)$. The central CTS spectrum is captured, crucial for isotope ratio sensing. For reactors, the spectral resolution can be reduced below 1 GHz.

Three THz PS systems with extant lasers were detailed, all having useful *SNR* for measurements of 10*cm* spatial resolution and good magnetic field resolution over ~1*ms*. Of note is the THz PS based on heterodyne receivers with electronic spectrometers, Gaussian beams and a diagnostic footprint as small as a CW laser of the same wavelength. Dual wavelength sources, QO systems and receivers allow a design canvassing all *key* ITS/CTS THz PS measurements. A ITS PS, $\lambda$<0.3*mm* is appropriate for sensing $n_e(s)$, $T_e(s)$, and $(\psi,\chi)(s)$; a CTS PS, $\lambda$>1*mm* is appropriate for sensing $n_q(s)$, $T_q(s)$, fast ions and confined alphas.

As a CTS diagnostic, heterodyne THz CTS PS match LP–HP gyrotron CTS performance when the *distributed* $N_m$ measurements are included. Continuous integration is possible at laser rep rates that inject THz pulses entering and leaving the plasma with only a brief pause between. CTS PS offers benefits beyond LP-HP CTS diagnostics, the *most important* of which is the ability to launch the beam in any direction within the port's FOV while maintaining maximal beam overlap regardless of beam refraction. Of importance: up to 100*x* brighter THz sources than LP–HP gyrotrons are possible and complete spectra. *Distributed* sensing with sightline steering provides 2D coverage of the poloidal plane on *sub-second* times and spatial coverage *on demand*(diagnostic agility) for *real-time* feedback in space as well as time for operational control and optimization.

NBI formed and sustained FRC plasmas were chosen to illustrate the utility of the THz CTS PS concept. The FRC has a promising equilibrium for aneutronic fusion that *still* lies beyond our current understanding. THz PS applied to FRCs would change a poorly diagnosed equilibrium to an admirably diagnosed equilibrium.

Diagnostic choices for fusion reactors are limited. While diagnosticians struggle with the cost and complexity of in-situ restoration of a first mirror surface, or combining multiple TS laser/detector systems to cover the wide ITS BW, THz diagnostics are naturally fusion reactor compatible. THz PS is manifestly an *advanced* diagnostic. In the THz range first mirror problems are mitigated, Schottky diodes and mixers provide exceptionally wide spectral BWs and broad spectral coverage, remote sensing *snap–shot* measurements of *key* equilibrium parameters provide *real-time* feedback, single port–small footprint and sightline steering are analogous to 2D imaging Doppler RADAR. *Advanced* $\mu$wave–THz diagnostics must contend with refraction, limiting beams to be {O–X} mode with $(\hat{s}\parallel\nabla n_e)$ which implies active steering in crossbeam geometries with integration time consumed by active feedback for optimizing alignment. These diagnostics are typically chord-averaged *non-local* measurements with noise contamination from the diagnostic's source. A THz Pulsed Spectropolarimeter replaces all limitations and handicaps with measurement accentuations: *local* measurements*(distributions)*, beam steering unimpaired



by refraction, one receiver for all sightline measurements and a contributed source noise that is *post*–measurement.

The inclusion of so many *key* parameters in one diagnostic is unprecedented. The measurements provide diagnostic *synergism* (*e.g.* measuring *local* $T_e$ provides correction for dispersion reliant measurements, $B_\parallel$ and $B_\perp$ and diagnostic *completeness*: *local B* and *local* particle kinetic pressures ($n_e T_e$, $n_e T_q$ and fast ions) of the charged plasma constituents are direct inputs to equilibrium solvers. The nature of LIDAR sensing provides measurement continuity in space ($\Delta s \sim 10 cm$) and time ($\tau_\text{plasma} \sim 1 ms$) which greatly aids in measurement accuracy and interpretation. The *figures of merit* for THz PS improve as peak $n_e$, $T_{q,e}$ and machine size increase on future fusion reactor devices.

The current laser and detector technology, together with the well-developed THz diagnostic expertise that exists within the plasma community provides a direct path to the realization of all three THz Pulsed Spectropolarimeter concepts on present day devices.

## ACKNOWLEDGEMENTS

The author would like to acknowledge engaging discussions on optical rectification and THz pulse generation techniques with Prof. Pietro Musumeci and Maximillian Lenz at UCLA. Also, helpful discussions regarding the vernier regenerative amplifier laser technique with Dr. Ashiq Fareed at TAE Technologies.



**APPENDIX A** Notation, terminology and simple relations

*Physical Constants:*

| | |
|---|---|
| $r_e, c, h$ | classical electron radius $2.8 \cdot 10^{-15}[m]$, speed of light $3 \cdot 10^8[m/s]$, Planck's constant $6.64 \cdot 10^{-34}[Js]$ |
| $k_b, \varepsilon_o, \mu_o$ | Boltzmann's constant, $1.38 \cdot 10^{-23}[°K]$, vacuum: permittivity, $8.85 \cdot 10^{-12}[F/m]$, permeability $4\pi \cdot 10^{-7}[H/m]$ |
| $M_p, m_e, e$, | Mass: hydrogen $1.67 \cdot 10^{-27}[Kg]$, electron $9.11 \cdot 10^{-31}[Kg]$, charge $1.6 \cdot 10^{-19}[C]$ |
| $C_{FR}, C_{CM}, C_n, C_B$ | Constants: FR effect $2.63 \cdot 10^{-13}[rad/T]$, CM effect $2.41 \cdot 10^{-11}[rad/m–T^2]$, density/field calibration constant |

*Acronyms:*

| | |
|---|---|
| LIDAR, RADAR, SOP, Yb, DSB | Light Detection and Ranging, Radio…, state of polarization, ytterbium, double side band |
| TS, ITS, CTS, LP-HP, LP | Thomson scattering: incoherent, collective, long pulse-high power, linearly polarized |
| PP, PS, CW, amu | Pulsed-(Polarimeter, Spectropolarimeter), continuous wave, atomic mass unit |
| SNR, LO, NBI, MHD | signal-to-noise ratio, local oscillator, neutral beam injection, magneto-hydrodynamic |
| MFE, FRC, SDF | magnetic fusion energy, field reversed configuration, spectral distribution function |
| RFP, MSE, DAQ, DD | reversed field pinch, motional Stark effect, data acquisition system, direct detection |
| DTL, ECE, ECA, ECRH | damage threshold limit, electron cyclotron-(emission, absorption, resonant heating) |
| DFG, TPF, FR, CM, FOV | difference frequency generation, tilted pulse front, Faraday, Cotton Mouton, field of view |
| QO, MP, EM, m.a., LN, THz, | quasi-optical, Martin-Puplett, electromagnetic, magnetic axis, lithium niobate, terahertz |
| OR, FM, RF, NIR, DC | optical rectification, figure of merit, radio frequency, near infra-red, direct current |
| FWHM, BW, GB, ES, BP | full width at half max, FWHM bandwidth, Gaussian beam, enhanced scattering, bandpass |
| NPS, PBS, RM, WG, ASE | (non-, ) polarizing beam splitter, roof mirror, wire grid, amplified spontaneous emission |

*Light pulse parameters:*

| | |
|---|---|
| $\lambda_o, f_o, (\psi,\chi), (\psi_d,\chi_d), (\psi_o,\chi_o)$ | wavelength$[m]$, frequency$[Hz]$, (polarization azimuth,ellipticity), detected, initial$[rad]$ |
| $\delta\nu_p, \tau_p, E_p, P_o(=E_p/\tau_p), \boldsymbol{E}$ | BW$[Hz]$, duration$[s]$, energy$[J]$, instantaneous power$[W]$, electric field$[V/m]$ |

*Scattering geometry and parameters:*

| | |
|---|---|
| $s, s_o(=s(0)), \hat{\boldsymbol{s}}, \Delta s, \tau_{\Delta s}(=2\Delta s/c)$ | Sightline: location along, direction, spatial resolution, $\Delta s$ resolving integration time, |
| $\alpha_o(s), \alpha_{sc}(s), \alpha_{tot}(=\alpha_o+\alpha_{sc})(s), \chi_{tot}(s)$ | Faraday Rotation: $s_o$ to $s$, scattered $s$ to $s_o$, detected, detected ellipticity, |
| $O_b, A_\mathcal{L}, A_\mathcal{R}$ | launch/receive beam overlap $[1/m]$, Beam areas: launch and receive at $V(s)$ |
| $(\boldsymbol{k}_i,\lambda_i), (\boldsymbol{k}_{sc},\lambda_{sc}), \boldsymbol{k}_s(=\boldsymbol{k}_{sc}-\boldsymbol{k}_i)$ | incident and scattered wave vectors and wavelengths and scattering wave vector$[1/m]$ |
| $\alpha_s(\theta_{sc})(=1/k_s\lambda_{De}), \theta_{sc}$ | Salpeter scattering parameter, scattering angle$[rad]$ |
| $V, A_d, w_o(=d/2), \Delta\Omega, \Delta\Omega_{f10}$ | Scattering: volume$[m^3]$, area$[m^2]$, radius$[m]$, solid angle$[sr]$, $\Delta\Omega$ in $0.008 sr$ |
| $M, \mathcal{E}, N_{ph}$ | # of EM modes collected, etendue$[m^2 sr]$(single mode–$\lambda_o^2$), # collected photons |
| $I_{p,s}(s)(=\int df \mathcal{I}_{s,p}(s,f)). \mathcal{I}_{s,p}(s,f), I_{tot}$ | 2 channel polarimeter Intensity$[W]$, spectral intensity$[W/Hz]$, total intensity$[W]$ |
| $S_{q,e,FI}(\hat{\boldsymbol{s}},\nu), S_{\{CTS,ITS,\delta E,\delta B\}}$ | $\hat{\boldsymbol{s}}$ directed SDF for species $q,e$ and confined fast ions, SDF{source of fluctuations}$[1/Hz]$ |

*Plasma parameters:*

| | |
|---|---|
| $\boldsymbol{B}, \boldsymbol{B}_{pol}, \boldsymbol{B}_\varphi, \boldsymbol{B}_{perp}, B_\|$ | Magnetic field$[T]$: total, poloidal plane, toroidal, perpendicular to $\hat{\boldsymbol{s}}$, parallel to $\hat{\boldsymbol{s}}$ |
| $B_\perp, \theta, \omega_{ce}(=e|\boldsymbol{B}|/m_e)$ | perpendicular to $\hat{\boldsymbol{s}}$ in poloidal plane, $\angle(\boldsymbol{B},\hat{\boldsymbol{s}})[rad]$, electron gyrofrequency$[rad/s]$ |
| $\lambda_{De}(=\sqrt{\varepsilon_o k_b T_e/e^2 n_e}), L_p, \beta_E, V_\phi$ | Debye length$[m]$, plasma size$[m]$, $\angle \boldsymbol{E},\boldsymbol{B}_{perp}[rad]$, ion azimuthal drift velocity$[m/s]$, |
| $\omega_{pe}(=\sqrt{e^2 n_e/\varepsilon_o m_e}), N_{\{\hat{e}_{\mu=1,2}\}}$ | plasma frequency$[rad/s]$, index of refraction for the two eigenpolarization states: $\{\hat{e}_{\mu=1,2}\}$ |
| $T_e, T_q, n_e, n_q, n_{FI}, Z_q$ | Temperature$[°K]$: electron, ion charge $q$, density: electron, ion, fast ion, ion charge state |
| $n_{20}, T_{q,e,10}, \Delta s_{10}, \lambda_{mm}, E_{400}$ | Units of: $n_e$ in $10^{20} m^{-3}$, $T_{q,e}$ in $10 keV$, $\Delta s$ in $10 cm$, $\lambda$ in $mm$, $E_p$ in $400 \mu J$ |

*Pulsed Spectropolarimeter instrument parameters:*

| | |
|---|---|
| $\Delta, N_{THz}, N_{NIR}, \delta(=\tau_p/(N_{THz}-1))$ | smallest crystal width$[m]$, # of THz pulses, # of NIR pulses, uniform pulse train delay$[s]$ |
| $E_{NIR}, \eta, \tau_D$ | NIR pulse energy$[J]$, conversion efficiency, delay between polarized pulse trains, |
| $R_p, N_s(R_p \cdot \tau_{plasma}), P_{NIR}(=R_p \cdot E_{NIR})$ | laser repetition rate$[Hz]$, number of scans in measurement time and NIR laser power$[W]$ |

*Detection parameters:*

| | |
|---|---|
| $P_n, T_n(=P_n/k_b\Delta f)$ | Noise power$[W]$ and mixer noise temperature$[°K]$ |
| $N_m(=L_p/\Delta s), N_s, T_p, \Delta T_p$ | # of *distributed* measurements, # scans in a measurement, time: of scan, between scans$[s]$ |
| $\delta\nu_{CTS, ITS, TS}, \delta\nu_T, \tau(=1/\delta\nu_T)$ | Spectral width: emission for CTS/ITS/TS$[Hz]$, detected$[Hz]$, detector integration time$[s]$ |
| $\delta f_{CTS}, \Delta f, E_{FI}$ | Downshifted $\delta\nu_{CTS}[Hz]$, electronic filter BW$[Hz]$, NBI beam energy$[J]$ |
| $P_{sc}, \Delta E_{sc}(=P_{sc}\tau_{\Delta s}), \wp_{sc}, \Delta \mathcal{E}_{sc}$ | Detected: power$[W]$, energy$[J]$, spectral: power$[W/Hz]$, energy$[J/Hz]$ |
| $f_{LO}, \tau_{DAQ}, \tau_{plasma}$, | LO frequency$[Hz]$, DAQ response time$[s]$, plasma dynamical integration time$[s]$ |

*FRC parameters:*

| | |
|---|---|
| $J_\varphi(r), \Psi(r), n_e(r), B_z(r)$ | azimuthal current density$[A/m^2]$, poloidal flux function$[Wb]$; midplane: density$[m^{-3}]$, axial B$[T]$ |



## APPENDIX B Coupled FR/CM effects: Mueller-Stokes' theory

A Pulsed polarimeter, optimally, has a large FR effect, $\alpha_{tot}(L_p) \sim 50\text{-}100°$ to spatially dissect $\alpha_{tot}$ with a well resolved $B_\parallel(s)$. If the CM effect is also large, $\alpha_{tot}(s)$ and $\chi_{tot}(s)$ may be coupled.

The magnetic field geometry is detailed in **Fig. B1**. The field $\boldsymbol{B}=\boldsymbol{B}_{pol}+\boldsymbol{B}_{tor}=\boldsymbol{B}_{pol}+B_{tor}\hat{\boldsymbol{\varphi}}$, $\hat{\boldsymbol{\varphi}}$ is the toroidal unit vector. The sightline, for this paper, lies in the poloidal plane, coords $(y,z)$. $\boldsymbol{B}_{pol}=\boldsymbol{B}_\perp+B_\parallel\hat{\boldsymbol{s}}=B_{pol,y}\hat{\boldsymbol{y}}+B_{pol,z}\hat{\boldsymbol{z}}$. The sightline coordinate system is $\{\hat{\boldsymbol{\varphi}}, \hat{\boldsymbol{s}} \times \hat{\boldsymbol{\varphi}}, \hat{\boldsymbol{s}}\}$. $B_\parallel = \boldsymbol{B} \cdot \hat{\boldsymbol{s}}$ with $\theta = \angle(\boldsymbol{B},\hat{\boldsymbol{s}})$. In the plane perpendicular to $\hat{\boldsymbol{s}}$ spanned by $(\hat{\boldsymbol{\varphi}}, \hat{\boldsymbol{s}} \times \hat{\boldsymbol{\varphi}})$, $\boldsymbol{B}_{perp}$ is defined with $\beta = \angle(\boldsymbol{B}_{perp}, \hat{\boldsymbol{s}} \times \hat{\boldsymbol{\varphi}})$. In a tokamak, $B_{tor}$ is a strong but for an FRC equilibrium, $B_{tor} \sim 0$ and then $\boldsymbol{B}_{perp} = \boldsymbol{B}_\perp$ with $\theta = \angle(\boldsymbol{B}_{pol}, \hat{\boldsymbol{s}})$ and $\beta = \{0, \pi\}$.

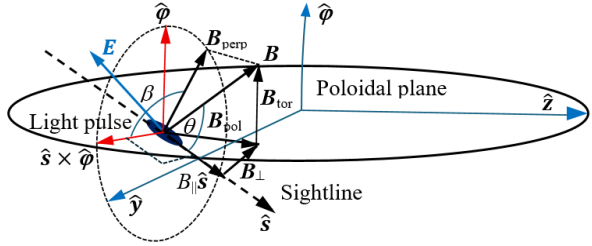

**Fig. B1.** A flux contour showing the light pulse trajectory in the poloidal plane, coords $\{\hat{\boldsymbol{\varphi}}, \hat{\boldsymbol{y}}, \hat{\boldsymbol{z}}\}$. $\boldsymbol{B}=B_{tor}\hat{\boldsymbol{\varphi}}+B_{pol,y}\hat{\boldsymbol{y}}+B_{pol,z}\hat{\boldsymbol{z}}$. $\boldsymbol{B}=\boldsymbol{B}_{perp}+B_\parallel\hat{\boldsymbol{s}}$ in the sightline coords $\{\hat{\boldsymbol{\varphi}}, \hat{\boldsymbol{s}} \times \hat{\boldsymbol{\varphi}}, \hat{\boldsymbol{s}}\}$, defining $\beta$. $\boldsymbol{E}$ encodes $(\psi, \chi)$.

The evolution of $(\psi, \chi)(s)$ with initial $(\psi_o, \chi_o)$ propagating in the magnetized plasma is elegantly elucidated in Segre[39]. The Stokes' vector intensities, $I_{S_i}$, $i=1,3$ given by (B1) and normalized intensities $S_i(=I_{S_i}/I_{pol}(s))$ determine $(\psi, \chi)(s)$. The $\alpha_{tot}(s)$ evolves according to the FR effect for $\boldsymbol{B}=B_\parallel\hat{\boldsymbol{s}}$ with $\Delta\chi=0$, see **Fig. B2**. A nonzero $\boldsymbol{B}_{perp}$ is responsible for producing a $\Delta\chi$ through the CM effect but can also produce a $\Delta\psi$ as well, making the physical interpretation of $\alpha_{tot}(s)$ as $B_\parallel(s)$ potentially ambiguous.

$$\vec{I}_S(\psi,\chi) = I_{pol}\hat{\boldsymbol{S}} = I_{pol} \begin{Bmatrix} \cos(2\chi)\cos(2\psi) \\ \cos(2\chi)\sin(2\psi) \\ \sin(2\chi) \end{Bmatrix} \quad \text{(B1)}$$

$$\psi = 1/2 \cdot \tan^{-1}(S_2/S_1); \quad 0 \leq \psi \leq \pi$$
$$\chi = 1/2 \cdot \sin^{-1}(S_3); \quad -\pi/4 < \chi \leq \pi/4$$

The degree of polarization, $D$ is given by $I_{pol}/I_{tot}$. Unpolarized emission (natural light) has $I_{pol}=0$ but a nonzero $I_{tot}(=\frac{1}{2}c\varepsilon_o|\boldsymbol{E}|^2)$. $D=1$ is assumed here.

The SOP as the pulse propagates to $s$ evolves as $\hat{\boldsymbol{S}}(s) = M_s\hat{\boldsymbol{S}}_o$. $M_s$ is a Mueller transformation, a 3x3 orthogonal matrix preserving the norm of $\hat{\boldsymbol{S}}$. For LIDAR, receiver $(\psi_d, \chi_d)$ is produced by a weak reflection(backscatter) at $s$ after which the SOP evolves in the reverse direction to $s_o$ at $t=2s/c$ with $B_\parallel$ reversed. The Mueller matrix for reflected coords, $(s \rightarrow -s)$ is $M_R = \{1,0,0;0,-1,0;0,0,-1\}$ with $M_R = M_R^{-1}$. Polarization is reflected in the $x$–$y$ plane. Under reflection, $\hat{\boldsymbol{S}}_R(s) = (M_R M_s M_R^{-1})(M_R\hat{\boldsymbol{S}}_o) = M_R M_s \hat{\boldsymbol{S}}_o$.

For $B(s)=B_\parallel$, a constant, $\alpha_o = C_{FR}\lambda_o^2 B_\parallel \int_0^s n_e ds'$ represented by $M_{-\alpha_o}=\{\cos(2\alpha_o),-\sin(2\alpha_o),0;\sin(2\alpha_o),\cos(2\alpha_o),0;0,0,1\}$. In retracing the path, $B_\parallel$ reverses sign producing $\hat{\boldsymbol{S}}_{LIDAR}(s) = M_{\alpha_o}M_R M_{-\alpha_o}\hat{\boldsymbol{S}}_o = M_{2\alpha_o}\hat{\boldsymbol{S}}_o$. $\alpha_{tot}(t=2s/c)=2\alpha_o$ expressing the *non*-reciprocity of the FR effect. On reflection, the eigenpolarizations, $\hat{e}_{+/-}$ are interchanged and with a reversed $B_\parallel$, $\alpha$ is doubled. A *distributed* $\alpha_{tot}(s)$ from a *pure* FR effect is characterized.

A Stokes' polarimeter consists of four intensity measurements, $I_{pol}\hat{\boldsymbol{S}}$ and $I_{tot}$, defining the 4 vector: $\Sigma = I_{pol}\{1,\hat{\boldsymbol{S}}\}$ for $D=1$. The $\Sigma$ intensities are resolved using polarizing elements at specific orientations[40]: $I_s = I_{0°,0°}$, $I_p = I_{90°,0°}$, $I_{45°,0°}$ and $I_{45°,90°}$. $\Sigma = \{I_{pol}(= I_s + I_p), I_s - I_p, 2I_{45°,0°} - \Sigma_o, 2I_{45°,90°} - \Sigma_o\}$.

For intuition about the evolution of the SOP under a general *distributed* magneto-optical activity, one appeals to the Poincaré sphere($S^2$) representation of $\hat{\boldsymbol{S}}(\psi,\chi)$ embedded in a 3-d space. The z-axis poles identify with $(\psi=0°, \chi=\pm45°)$ and $\hat{\boldsymbol{S}}=\{0,0,\pm1\}$; x-axis poles: $(\psi=\{0°,90°\},\chi=0°)$ and $\hat{\boldsymbol{S}}=\{\pm1,0,0\}$; y-axis poles $(\psi=\pm45°,\chi=0°)$ and $\hat{\boldsymbol{S}}=\{0,\pm1,0\}$ as illustrated in **Fig. B2**. Orthogonal states identify as antipodal points. Mueller matrices are $SU(2)$ group transformations. An $SU(2)$ Gadget[41] is a realization of a polarization state generator mapping $(\psi_o, \chi_o)$ into any $(\psi, \chi)$ using two ¼– and one ½– waveplates for a given $\lambda$.

**Fig. B2.** A Mueller transformation of SOP $(\pi/4,0)$ to $(\pi/2,\chi)$ is shown as a rotation about an $\vec{\Omega}$ axis. If $\vec{\Omega} \not\parallel S_3$, then $\Delta\chi \neq 0$. For $\vec{\Omega} \parallel S_3$ a pure Faraday rotation results. 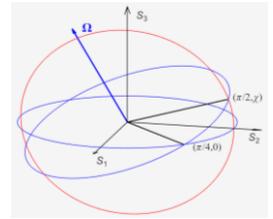

The incremental change $(\Delta\psi, \Delta\chi)$ over $ds$ is given by Eq (B2),

$$d\hat{\boldsymbol{S}}(s)/ds = \vec{\Omega}(s) \times \hat{\boldsymbol{S}}(s) \quad \text{(B2)}$$

where $\vec{\Omega}(s)$ is given in Eq (B3), a rotation vector, aligned with the antipodal eigenpolarizations of the medium at $s$: $\hat{\boldsymbol{s}}_{s,f}(s)$ with slow $N_s$ and fast $N_f$. $\vec{\Omega}[rad/m] = \Delta\varphi/\Delta s\,\hat{\boldsymbol{s}}_{s,f} = k(N_s - N_f)\hat{\boldsymbol{s}}_{s,f}$ for rotation $\Delta\varphi$. In the coord system[39], $\vec{\Omega}(s)$ is given by Eq (B3) for $\omega \gg \omega_{ce}$.



$$\vec{\Omega}(\lambda, n_e, \boldsymbol{B}, s) \approx \begin{cases} 2C_{CM}\lambda^3 n_e(B_\perp^2 - B_\varphi^2) \\ 2C_{CM}\lambda^3 n_e(2B_\perp B_\varphi) \\ 2C_{FR}\lambda^2 n_e B_\parallel \end{cases} \quad \text{(B3)}$$

When the activities are so weak that they are independent (*uncoupled*), the evolution equation simplifies to three line integrated quantities: $W_j(s) = \int_0^s ds'\, \Omega_j(s')$; $\boldsymbol{S}(s) = W_j(s) \times \hat{\boldsymbol{S}}_o + \hat{\boldsymbol{S}}_o$, leading to the cold dispersion line integrated formulas.

The algorithm for LIDAR $(\psi_d, \chi_d)(t=2s/c)$ sourced from $s$ is found by iterating Eq (B2): $\hat{\boldsymbol{S}}(s_k) = \langle \Omega \rangle_{k-1} \times \hat{\boldsymbol{S}}(s_{k-1})\Delta s + \hat{\boldsymbol{S}}(s_{k-1})$, $\langle \vec{\Omega} \rangle_{k-1} = \left(\vec{\Omega}(s_k) + \vec{\Omega}(s_{k-1})\right)/2$; $\hat{\boldsymbol{S}}(s_k)$ scaled to unit norm, with initial $\hat{\boldsymbol{S}}_o = \hat{\boldsymbol{S}}(s=0)$. At $s$, $\hat{\boldsymbol{S}}(s) = M_{\vec{\Omega}(s)}\hat{S}_o$. The reflection matrix, $M_R$ is applied to $M_{\vec{\Omega}(s)}\hat{S}_o$ to transform to reflected coords, $\hat{\boldsymbol{S}}_R(s) = (M_R M_{\vec{\Omega}(s)} M_R^{-1}) M_R \hat{\boldsymbol{S}}_o = M_R M_{\vec{\Omega}(s)}\hat{\boldsymbol{S}}_o$. The SOP then evolves from $s$ to $s_o$, using $\vec{\Omega}(s)$ with $B_\parallel \rightarrow -B_\parallel$.

A 4 ch polarimeter measures $\Sigma = \{I_{tot}, I_{pol}\hat{\boldsymbol{S}}\}(t)$, a 2 ch. polarimeter measures only $I_{s,p}(s)$, and derives $(\psi(s), I_{pol}(s))$ assuming $\chi(s)$ is neglectable. The normalized intensities are given by Eq (B4) with $I_{pol}=\tfrac{1}{2}c\varepsilon_0|\boldsymbol{E}|^2 = I_s + I_p$,

$$\begin{aligned} S_1 &= (I_s - I_p)/I_{pol} \\ S_2 &= (I_{45°} - I_{-45°})/I_{pol} \\ S_3 &= (I_+ - I_-)/I_{pol} \end{aligned} \quad \text{(B4)}$$

Returning to **Fig. 6b**, the coupled theory for strong CM and FR effects is shown in **Fig. B3** for this data set.

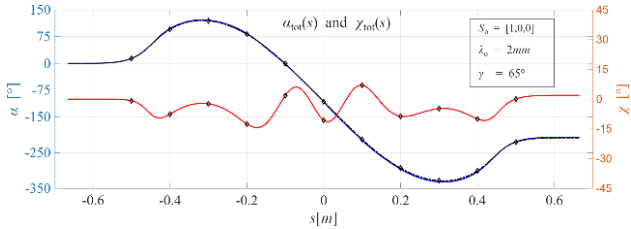

**Fig. B3.** Strong FR/CM effects when $\lambda_o$ is a long, 2*mm*. The $\alpha(s)$ trace is also plotted as a line integrated trace(dotted) and as a coupled FR/CM trace(solid), indistinguishable. The CM trace is much reduced compared to the line integrated CM trace of **Fig. 6b**.

The coupled CM effect is small because the linear retardance depends on the orientation of **E** relative to $\boldsymbol{B}_\perp$ and with an overwhelmingly strong FR effect, **E** rotates rapidly subduing the CM effect. This, experimentally, suggests that a 2 ch polarimeter will suffice even with $\lambda_o = 2mm$.

When the FR effect is weakened by increasing $\gamma$ to 85° the traces are shown to be strongly coupled and the coupled FR trace no longer overlays the line integrated FR trace, **Fig. B4**. The glitch in the $\alpha_{tot}(s)$ trace is superficial, a trig singularity.

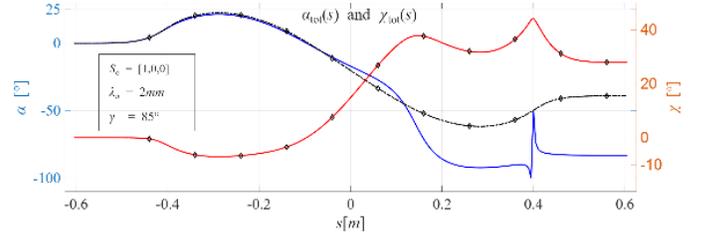

**Fig. B4.** The FR effect is weakened relative to the CM effect as $\gamma$ is increased from 65° to 85°, $B_\perp/B_\parallel(s)=tan\gamma=11.4$. $\alpha_{tot}$ deviates from the line integrated $\alpha_{tot}$ trace as $\Delta\chi_{tot}$ approaches 45°.

In this case, a general 4 channel time division Stokes' vector polarimeter would be needed to analyze the $(\psi,\chi)(t)$ traces and recover $(B_\parallel, B_\perp)(s)$ from the SOP history which is shown in **Fig. B5** on the Poincaré sphere for the traces in **Fig. B4**.

**Fig. B5.** The $\chi_{tot}(s)$, black trace, from **Fig. B4** is plotted on the Poincare sphere. The pathology in the $\alpha_{tot}(s)$ trace is seen to occur when $\chi_{tot}(s)$ approaches the upper pole where $\psi$ becomes ambiguous.

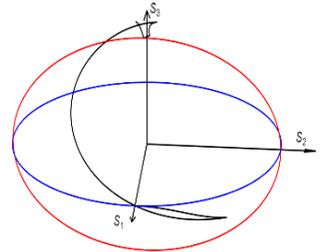

For the mid-plane FRC profile measurements a 2 channel polarimeter is shown to suffice with a judicious choice of the sightline $\gamma$ on FRC equilibrium. The ratio of FR:CM strengths vary with $\gamma$ as well as with $\lambda_o$.